\documentclass[12pt,a4paper]{article}
\pdfoutput=1 % 
\usepackage[utf8]{inputenc}
\usepackage{fullpage}
\usepackage{cite}
\usepackage{subcaption}
\usepackage[english]{babel}
\usepackage{amsmath}
\usepackage{physics}
\usepackage{amsthm}
\usepackage{amsfonts}
\usepackage{amssymb}
\usepackage{bbold}
\usepackage{mathrsfs}
\usepackage{mathbbol}
\usepackage[toc,page]{appendix}
\usepackage{subcaption}
\usepackage{bbding}
\usepackage{esvect}
\usepackage{commath}
\usepackage{enumitem}
\usepackage[affil-it]{authblk}
\usepackage{siunitx}
\usepackage{braket}
\usepackage{phaistos}
\usepackage{multicol}
\usepackage{overpic}
\usepackage{comment}
\usepackage[colorlinks=true
  ,urlcolor=blue
  ,anchorcolor=blue
  ,citecolor=blue
  ,filecolor=blue
  ,linkcolor=blue
  ,menucolor=blue
  ,pagecolor=blue
  ,linktocpage=true
  ,pdfproducer=medialab
  ,pdfa=true
]{hyperref}
\usepackage{xcolor}

\usepackage{jheppub}
\usepackage{tikz}
\usetikzlibrary{decorations.markings, arrows.meta, calc}
\usepackage{graphicx}
\usepackage{quantikz}

\usepackage{environ}
\NewEnviron{myequation}{%
\begin{equation*}
\scalebox{1.7}{$\BODY$}
\end{equation*}
}

\newtheorem{theorem}{Theorem}
\newtheorem{corollary}[theorem]{Corollary}

\newtheorem{conjecture}{Conjecture}

\usepackage{bbm}% Added by cindy for that nice \mathbbm{1}

\newcommand{\Hil}{\mathcal{H}}

\newcommand{\etaUnionSum}{\sum_{j=1}^n Z_j(-\eta_j \cup_{\partial_{\eta_j}} \eta_j)}

\def\ket#1{{|{#1}\rangle}} 

\title{Topological Preparation of Non-Stabilizer States and Clifford Evolution in $SU(2)_1$ Chern-Simons Theory}

\author[a,b]{William Munizzi,}
\author[c]{Howard J. Schnitzer}

\affiliation[a]{Division of Physical Sciences, College of Letters and Science, University of California, Los Angeles, Los Angeles, CA, 90095, USA}
\affiliation[b]{Department of Physics, Arizona State University, Tempe, AZ 85281, USA}
\affiliation[c]{Martin Fisher School of Physics, Brandeis University, Waltham, MA 02453, USA}

\emailAdd{wmunizzi17@ucla.edu}
\emailAdd{schnitzr@brandeis.edu}

\abstract{We develop a topological framework for preparing families of non-stabilizer states, and computing their entanglement entropies, in $SU(2)_1$ Chern-Simons theory. Using the Kac-Moody algebra, we construct Pauli and Clifford operators as path integrals over $3$-manifolds with Wilson loop insertions, enabling an explicit topological realization of $W_n$ and Dicke states, as well as their entanglement properties. For the genus-$1$ case, the modular $SL(2,\mathbb{Z})$ generators on a single torus boundary realize the standard Clifford generators. We further conjecture a higher-genus extension of this correspondence specifically for the Clifford orbits of $W$ and Dicke states, in which the orbit of $W_n$ is represented by a sum over genus-$g$ handlebodies with Dehn twists applied to each torus. Our results extend the known correspondence between Clifford structures and modular transformations, and demonstrate that non-stabilizer states can be understood as controlled deformations within the $SU(2)_1$ modular tensor category, providing a concrete example of how non-stabilizer quantum resources emerge from purely topological data.}

\begin{document} 
\maketitle
\flushbottom

%%%%%%%%%%%%%%%%%%%%%%%%%%%%%%%%%%%%%%%%%%%%%%%%%%%%%%%%%%%%%%%%%%%%%%%%%%%%%%%%%%%%%%%

\section{Introduction}

Entanglement remains one of the most profound features of quantum mechanics, underlying a wide range of phenomena and driving many significant advances in contemporary quantum science and technology. Given this central importance, efforts to understand and characterize the structure of entanglement in many-body quantum systems have developed into a vibrant area of ongoing research. Quantitative measures such as entanglement monotones, e.g. entanglement entropy, provide a functional description of the non-classical correlations shared among subsystems of multipartite quantum states. In topological quantum field theory (TQFT), multipartite entanglement measures are constrained by the geometric and topological properties of the surfaces on which the theory is defined. The manner in which topology encodes precise features of entanglement reveals deep connections between geometry and information, with implications ranging from topological order in quantum materials to emergent spacetime features in AdS/CFT.

Chern-Simons theory, particularly in three dimensions, likewise exhibits strong connections to the topological structure of quantum systems. A prominent example arises in knot theory, where Witten's seminal work~\cite{Witten:1998qj} demonstrated that the expectation values of Wilson loops in Chern-Simons theory reproduce notable knot invariants, e.g. the Jones polynomial, by way of skein relations. This correspondence emphasized a profound link between TQFT and knot theory, inspiring research efforts connecting topological features, e.g. link polynomials and braid group representations, to physical properties of quantum field theories~\cite{Balasubramanian_2017,Balasubramanian_2018,Balasubramanian_2025}. A related approach defines certain properties of a Chern-Simons theory in terms topological manifolds. Foundational work in this direction was initiated for stabilizer quantum states~\cite{Salton:2016qpp}, although relatively little progress has been made in extending these ideas to more general quantum states. 

A central aspect in the topological realization of quantum states and their entanglement entropies relies on the construction of Chern-Simons quantities as three-dimensional torus amplitudes~\cite{Salton:2016qpp}. In this work, we revisit the representation of amplitudes, entanglement entropies, and Clifford orbits within an $SU(2)$ Chern-Simons framework. In particular, we consider an $SU(2)_1$ Chern-Simons realization of $W_n$ states and their properties, which we relate to the topological features of state preparation on a manifold.

Beyond providing a topological preparation for states, we examine the dynamics generated by the Clifford group acting on states. This perspective establishes a correspondence between the action of quantum operators, and the associated geometric transformations of the underlying manifold. Throughout the different applications considered, we consistently encounter sums over products of $3$-dimensional handlebodies, constructed using Heegaard splittings of manifolds subject to transformations of the mapping class group. Each mapping class group is generated by the modular transformation matrix $S$, and a Dehn twist generated by fusion matrix $T$, where $S$ and $T$ are obtained from the Kac-Moody algebra $SU(2)_1$.

In this paper we consider state preparation for topological field theories in an $SU(2)_1$ Chern-Simons formalism. We use our construction to algebraically compute the bipartite entanglement entropy for $W_n$ states, a class of states integral to quantum computing, and characterize the dynamics of entanglement under Clifford orbits. In Section \ref{Background} we provide a review of entanglement calculations in Chern-Simons theory, and provide an algebraic construction for Pauli and Clifford operators in the $SU(2)_1$ Kac-Moody algebra. In Section \ref{WEntropies} we explicitly describe how to prepare $W_n$ states and construct their entanglement entropies. Section \ref{CliffordKacMoody} extends the group theoretic construction of Clifford orbits to a topological formalism, which we then employ to analyze the evolution of entanglement under this set of quantum operators. We propose several conjectures for manifolds, and their associated topological transformations, which realize the algebra of Clifford evolution. We propose an extension of our $W_n$ state protocol to the full set of Dicke states, of which $W_n$ is a special case. We conclude with a discussion on generalizations and applications of this work.

\section{Review}\label{Background}

In this section we offer a background review of necessary foundational material regarding entanglement calculations in Chern-Simons theory, qudit quantum groups, and algebraic structures in $SU(2)$. Readers familiar with these topics should feel free to skip this review material.

\subsection{Entanglement Entropy in Chern-Simons Theory}\label{ChernSimonsEntropySection}

Entanglement entropy serves as a fundamental measure for quantifying non-classical correlations in quantum systems, specifically those shared between subsystems of a bipartite quantum state. Extensive investigations of entanglement entropy have led to classifications on the space of quantum states~\cite{Linden:2013kal,Bao2015, Bao2020,Keeler:2022ajf, Keeler:2023xcx, Munizzi:2023ihc,Bao2020a,Bao2022a,Fuentes:2025kwg,Munizzi:2024huw}, and established constraints for information processing~\cite{Hayden2013,Ingleton1971,Khumalo:2025xfv} and entanglement dynamics~\cite{Keeler:2023shl}. Throughout this work we distinguish between the local single-qudit Hilbert space $\Hil_{\rm loc}$, of dimension $d \equiv \dim \Hil_{\rm loc}$, and the multipartite Hilbert space $\Hil = \Hil_{\rm loc}^{\otimes n}$, of dimension $d^n$, on which an $n$-party state $\psi$ is defined. For an $n$-party quantum state $\psi \in \Hil$, with the factorization $\Hil = \Hil_{\rm loc}^{\otimes n}$, let $I$ denote an $\ell$-qudit subsystem of $\psi$. The entanglement entropy of $I$, with respect to its $(n-\ell)$-qudit complement $\overline{I}$, is computed using the von Neumann entropy
\begin{equation}\label{SubsystemEntropyDefinition}
   S_{I} = -\Tr \left(\rho_{I} \log_d \left(\rho_{I}\right) \right),
\end{equation}
where the log base $d = \dim \Hil_{\rm loc}$ is the local Hilbert space dimension%
\footnote{The log base is set by the units of information, e.g. \textit{bits} for $d=2$, or \textit{dits} more generally.} %
and $\rho_I = \Tr_{\overline{I}} \ket{\psi}\bra{\psi}$ is the reduced density matrix of $I$, built by tracing over all subsystems in $\psi$ which are not members of $I$.

Chern-Simons theory is a topological quantum field theory~\cite{Witten1989,Elitzur1989,MooreSeiberg1989,Atiyah1990} in three dimensions, defined by the Chern-Simons action
\begin{equation}\label{CSaction}
    S_{CS} = \frac{k}{4\pi}\int_{\mathcal{M}} \Tr \left(A \wedge dA + \frac{2}{3}A \wedge A \wedge A\right),
\end{equation}
where $\mathcal{M}$ is an oriented $3$-manifold, and $A \in SU(2)$ is a connection one-form that serves the role of the gauge field. One natural choice of gauge-invariant observables is the set of Wilson loops around an oriented closed curve $C \in \mathcal{M}$, given by
\begin{equation}\label{WilsonLoop}
    W(C,R) = \Tr_R\left[\mathcal{P} \exp\left(i \oint_C A\right)\right],
\end{equation}
with $R$ an irreducible representation for $SU(2)$, and $\mathcal{P}$ a path-ordering on $C$.

For Wilson loop operators with support on a link $L \equiv \bigcup_{i=1}^n C_i$, the corresponding observable $W(L,R)$ takes the form of the product
\begin{equation}\label{WilsonLoopUnion}
    W(L,R) = W(C_1,R_1)...W(C_n,R_n),
\end{equation}
with each $R_i$ an irreducible representation of $SU(2)$ associated to $C_i$. When the links $L$ may be thickened to solid tori~\cite{Chun:2017hja}, the expectation value of $W(L,R)$ is well-defined
\begin{equation}\label{WilsonLoopFraction}
    \langle W(L,R) \rangle_{\mathcal{M}} = \frac{Z(\mathcal{M};L,R)}{Z(\mathcal{M})},
\end{equation}
where $Z(\mathcal{M}) \equiv \int \mathcal{D}A 
 e^{iS_{CS}}$ gives the partition function on $\mathcal{M}$, such that
\begin{equation}\label{PartitionFunction}
    Z(\mathcal{M};L,R) = \int \mathcal{D}A 
 e^{iS_{CS}} W(L,R).
\end{equation}
For a many-torus subsystem $A \subseteq \partial \mathcal{M}$, when the boundary state prepared by the path integral has flat entanglement spectrum (as is the case for stabilizer states prepared in $U(1)_k$ Chern-Simons theory at odd prime level~\cite{Salton:2016qpp}), the entanglement entropy $S(A)$ admits a single-replica expression~\cite{Salton:2016qpp,Chun:2017hja}
\begin{equation}\label{TopologicalEE}
    S(A) = -\log\left[\frac{Z(-2\mathcal{M} \cup_{f_A} 2\mathcal{M})}{Z(-\mathcal{M} \cup_{\partial \mathcal{M}} \mathcal{M})^2}\right].
\end{equation}
The object $2 \mathcal{M} = \mathcal{M} \cup \mathcal{M}$ represents the two copies of manifold $\mathcal{M}$ used to prepare the pure state, while $f_A$ denotes an exchange diffeomorphism on the tori in $A$ which leave the boundary subset $\partial M \backslash A$ invariant. The right-hand side of Eq.~\eqref{TopologicalEE} computes the second R\'enyi entropy $S_2(A)$, which equals the von Neumann entropy $S(A)$ when the entanglement spectrum is flat~\cite{Salton:2016qpp}. For generic states prepared by the path integral, including the non-stabilizer $W_n$ and Dicke states considered in this work, the entanglement spectrum need not be flat, and Eq.~\eqref{TopologicalEE} returns $S_2(A)$ rather than $S(A)$. In Section~\ref{WEntropySection} we therefore do not use Eq.~\eqref{TopologicalEE} to compute the von Neumann entropy of $W_n$. We instead evaluate $S(A)$ directly from the reduced density matrix and then provide a topological re-expression of the resulting closed-form scalar in terms of partition functions over copies of $\eta$.

\subsection{The Qudit Pauli Group}

The Pauli matrices~\cite{NielsenChuang2010} are a set of unitary and Hermitian operators represented, in the two-dimensional computational basis $\{\ket{0}, \ket{1}\}$, by the following matrices
\begin{equation}\label{PauliMatrices}
\mathbb{1} = \begin{bmatrix}1&0\\0&1\end{bmatrix}, \,\, \sigma_X = \begin{bmatrix}0&1\\1&0\end{bmatrix}, \,\,
    \sigma_Y = \begin{bmatrix}0&-i\\i&0\end{bmatrix}, \,\,
    \sigma_Z = \begin{bmatrix}1&0\\0&-1\end{bmatrix}.
\end{equation}
The matrices in Eq.\ \eqref{PauliMatrices} form a group under multiplication~\cite{Gottesman1997,Gottesman1998}, known as the single-qubit Pauli group $\Pi_1$
\begin{equation}
    \Pi_1 = \langle \sigma_X, \sigma_Y, \sigma_Z \rangle.
\end{equation}
For the two-dimensional Hilbert space $\Hil = \mathbb{C}^2$, the action of $\Pi_1$ generates the algebra of linear operations $L(\Hil)$.

Pauli group action on a multiqubit system is defined as a product of operators from $\Pi_1$, each of which acts locally on a single qubit. For example, the action of $\sigma_X$ on the $k^{th}$ qubit in an $n$-qubit system is performed by
\begin{equation}\label{PauliString}
    I^1\otimes\ldots\otimes I^{k-1} \otimes \sigma_X^k \otimes I^{k+1} \otimes \ldots \otimes I^n.
\end{equation}
Composite operations as in Eq.\ \eqref{PauliString} are called Pauli strings, with the weight of each string indicated the number of non-identity operators. The $n$-qubit Pauli group $\Pi_n$ is multiplicative group generated by all Pauli strings of weight $1$.

We can naturally extend~\cite{Gottesman1999} the action of $\Pi_n$ to Hilbert spaces of dimension $d>2$ by constructing the following $d$-dimensional generalizations of $\sigma_X$ and $\sigma_Z$
\begin{equation}\label{PauliQuditOperators}
    X = \sum_{x \in \mathbb{Z}_d}\ket{x+1}\bra{x}, \quad Z = \sum_{z \in \mathbb{Z}_d} \omega^z\ket{z}\bra{z},
\end{equation}
where the factor $\omega \equiv \exp(2\pi i/d)$ is the $d^{th}$ root of unity. Eq.\ \eqref{PauliQuditOperators} returns the correct dimensionality for odd dimension $d$; however, when $d$ is even the operator $XZ$ has order $2d$, and therefore generates additional roots of unity. Accordingly, the convention $\hat{\omega} \equiv \exp(2\pi i/D)$, with
\begin{equation}
D = 
\begin{cases} 
    d, &  \textnormal{d odd}\\
    2d, &  \textnormal{d even}\\
\end{cases}
\end{equation}
is often employed. The set of all operators $\hat{\omega}^rX^aZ^b$, for $r\in \mathbb{Z}_D$ and $a,b \in \mathbb{Z}_d$, defines the single-qudit Pauli group $P^d$.

As with systems of qubits, we extend the action of $P^d$ to systems of multiple qudits by constructing operators $X^aZ^b$ as the tensor product
\begin{equation}\label{PauliQuditStrings}
    X^\textbf{a}Z^\textbf{b} = X^{a_1}Z^{b_1} \otimes ... \otimes X^{a_n}Z^{b_n},
\end{equation}
with commutation relation
\begin{equation}\label{PauliQuditCommutation}
\left(X^a Z^b \right)\left(X^{\overline{a}}Z^{\overline{b}} \right) =\omega^{\sum_{i=1}^n a_i \overline{b}_i - \overline{a}_i b_i}\left(X^{\overline{a}}Z^{\overline{b}}\right) \left(X^a Z^b \right).
\end{equation}
The group generated by operations as in Eq.\ \eqref{PauliQuditStrings} builds the $n$-qudit Pauli group $P^d_n$, up to global phase.

\subsection{The Qudit Clifford Group}\label{QuditCliffordSection}

The Clifford operators comprise a well-known set of quantum operations~\cite{NielsenChuang2010}, famous for their ability to be efficiently simulated on a classical computer~\cite{aaronson2004improved,Gottesman1997}. Furthermore, the well-known set of stabilizer quantum states can be generated using only the action of Clifford operations on measurement basis states. In two dimensions, the Clifford operators form a multiplicative matrix group $\mathcal{C}_n$ generated by the Hadamard, phase, and $CNOT$ quantum gates
\begin{equation}\label{CliffordGates}
    H\equiv \frac{1}{\sqrt{2}}\begin{bmatrix}1&1\\1&-1\end{bmatrix}, \quad P\equiv \begin{bmatrix}1&0\\0&i\end{bmatrix}, \quad     C_{i,j} \equiv \begin{bmatrix}
            1 & 0 & 0 & 0\\
            0 & 1 & 0 & 0\\
	    0 & 0 & 0 & 1\\
	    0 & 0 & 1 & 0
            \end{bmatrix}.
\end{equation}
The group $\mathcal{C}_n$ is the unitary normalizer of $\Pi_n$, mapping the Pauli group to itself under conjugation.

We can extend the Clifford group action to arbitrary dimension~\cite{Gross2006,Hostens2005} by constructing the $d$-dimensional analog of the Hadamard, phase, and $CNOT$ operations in Eq.\ \eqref{CliffordGates}. The $d$-dimensional generalization of the Hadamard gate is the modular transformation
\begin{equation}\label{GeneralizedHadamard}
    S^* = \frac{1}{\sqrt{d}} \sum_{a=0}^{d-1}\sum_{b=0}^{d-1}\omega^{ab} \ket{a}\bra{b},
\end{equation}
which initializes a transformation on the Pauli operators as
\begin{equation}\label{QFTAction}
\begin{split}
    X &\mapsto Z,\\
    Z &\mapsto X^{-1}.\\
\end{split}
\end{equation}
We can rewrite Eq.\ \eqref{GeneralizedHadamard}, in a more familiar form, as the discrete $d$-dimensional quantum Fourier transform $S$ on the basis state $\ket{a}$
\begin{equation}\label{QFT}
    S\ket{a} = \frac{1}{\sqrt{d}} \sum_{b=0}^{d-1}\omega^{ab} \ket{b}.
\end{equation}

When extending the phase gate in Eq.\ \eqref{CliffordGates} to $d>2$ dimensions, we consider the operator which maps
\begin{equation}\label{PhaseAction}
\begin{split}
    X &\mapsto XZ,\\
    Z &\mapsto Z.\\
\end{split}
\end{equation}
A unitary representation that transforms Pauli operators as in Eq.\ \eqref{PhaseAction} depends on the even/odd parity of Hilbert space dimension $d$, and can be expressed as the action on $\ket{j} \in \Hil^d$ as
\begin{equation}\label{GeneralizedPhase}
\begin{split}
    \overline{P}\ket{j} &= \omega^{\frac{j(j-1)}{2}}\ket{j}, \quad \textnormal{d odd},\\
    \overline{P}\ket{j} &= \omega^{\frac{j^2}{2}}\ket{j}, \quad \textnormal{d even}.\\
\end{split}
\end{equation}
Together, the operators $S$ and $\overline{P}$ are a necessary and sufficient set~\cite{Farinholt_2014} to generate the single-qudit Clifford group.

Extending the Clifford group to multi-qudit action~\cite{Farinholt_2014} requires the addition of the conditional-sum%
\footnote{The $C_{sum}$ gate is sometimes referred to as the controlled-addition $C_{add}$ gate in the literature.}
gate $C_{sum}$, the generalization of $CNOT$ to $d$-dimensional systems, which initiates the following map
\begin{equation}\label{CSumMapping}
  \begin{split}
      X \otimes I &\mapsto X \otimes X,\\
      I \otimes X &\mapsto I \otimes X,\\
      Z \otimes I &\mapsto Z \otimes I,\\
      I \otimes Z &\mapsto Z^{-1} \otimes Z.\\
  \end{split}  
\end{equation}
The $C_{sum}$ operator acts on a qudit pair $\{i,j\}$, in an $n$-qudit system, with the action
\begin{equation}\label{CSum}
    C_{sum}\ket{i}\ket{j} = \ket{i}\ket{i+j \pmod{d}}.
\end{equation}
where $d$ indicates the Hilbert space dimension, and $i,j$ denote the control and target qubits, respectively. As an example, the action of $C_{sum}$ on a pair of qutrits can be written in the symplectic matrix form
\begin{equation}\label{CSumMatrix}
    C_{sum_{1,2}} = \begin{bmatrix}
1 & 0 & 0 & 0 & 0 & 0 & 0 & 0 & 0 \\
0 & 1 & 0 & 0 & 0 & 0 & 0 & 0 & 0 \\
0 & 0 & 1 & 0 & 0 & 0 & 0 & 0 & 0 \\
0 & 0 & 0 & 0 & 0 & 1 & 0 & 0 & 0 \\
0 & 0 & 0 & 1 & 0 & 0 & 0 & 0 & 0 \\
0 & 0 & 0 & 0 & 1 & 0 & 0 & 0 & 0 \\
0 & 0 & 0 & 0 & 0 & 0 & 0 & 1 & 0 \\
0 & 0 & 0 & 0 & 0 & 0 & 0 & 0 & 1 \\
0 & 0 & 0 & 0 & 0 & 0 & 1 & 0 & 0 \\
\end{bmatrix}.
\end{equation}

The unitary representation of $C_{sum}$ given by Eq.\ \eqref{CSum} is sufficient to preserve the map in Eq.\ \eqref{CSumMapping}, up to global phase, for arbitrary $d$. When global phase cannot be neglected, however, the following modification is required to account for even dimension
\begin{equation}\label{CSumEven}
    C_{sum}\ket{i}\ket{j} = \omega^{\frac{1}{2}(i+j)}n\ket{i}\ket{i+j, \pmod{d}}.
\end{equation}
Together, the discrete quantum Fourier transform, phase, and controlled-sum operators generate the $n$-qudit Clifford group.

\subsection{Kac-Moody Algebra and $SU(2)_1$ Pauli Group}\label{KacMoodyPauli}

A Kac-Moody algebra~\cite{Kac1990,PressleySegal1986} is the generalization of a Lie algebra, particularly useful for describing symmetries of infinite-dimensional physical systems, as well as scale-invariant field theories. One specific realization in contemporary physics surrounds the Kac-Moody construction of $SU(2)$, denoted $SU(2)_k$ where $k$ is the level (or central extension) of the theory. The Kac-Moody extension of $SU(2)$, to a single level ($k=1$), describes the conformal field theory~\cite{Goddard1976,Witten1984,KnizhnikZamolodchikov1984,Verlinde1988} for a single free boson with central charge $c=1$.

The group $SU(2)$ admits a single Casimir operator, enabling all its irreducible representations to be expressed using a Young Tableaux consisting of a single column, with each element corresponding to an index in the fundamental representation. The algebra $SU(2)_1$ is further constrained, restricting all representations to the level-$1$ cutoff. Accordingly, the Young Tableaux for $SU(2)_1$ consists of a single column with two elements, which we index as $a=0,\,1$. The fusion of two integrable representations of $SU(2)_1$ is encoded by the fusion tensor $N^{a_3}_{a_1, a_2}$, whose components count the number of independent fusion channels from $a_1 \otimes a_2$ to $a_3$. Fixing one input index to the fundamental representation $a_2 = 1$ (corresponding to a Wilson loop in the spin-1/2 representation, see Section~\ref{TopologicalPreparationSection}) and reading $N$ as a map from $\Hil_{\rm loc} \to \Hil_{\rm loc}$ defined by its matrix elements, we have~\cite{Schnitzer:2020tiv}
\begin{equation}\label{FusionMatrix}
    \bra{a_3} N \ket{a_1} \;\equiv\; N^{a_3}_{a_1, 1} \;=\; \delta_{a_3,\, a_1 + 1\, (\!\!\!\!\!\!\mod 2)},
\end{equation}
which reproduces the action $X \ket{a} = \ket{a+1\, (\!\!\!\!\mod 2)}$ of the Pauli $X$ operator. In this way, the fusion tensor in Eq.\ \eqref{FusionMatrix} provides an algebraic realization for the Pauli $X$ operation, with the row and column indices identified with the input and output qudit states.

As described in Eq.\ \eqref{QFTAction}, a Hadamard operation $S_{ab}$ can be used to conjugate the Pauli $X$ operator to produce a Pauli $Z$ operation, i.e.\ $Z = S X S^\dagger$. Employing this relation with Eq.\ \eqref{FusionMatrix}, we can give an algebraic expression for the matrix elements of Pauli $Z$ as
\begin{equation}\label{KacPauliZ}
    Z_{a a'} \;=\;  \sum_{b,\, c\, =\, 0}^{1} S_{a b}\, N^{c}_{b, 1}\, (S^\dagger)_{c\, a'}, 
\end{equation}
where $S_{ab}$ is the modular transformation matrix for $SU(2)_1$. In the standard normalization, when $a,\,b = 0,\,1$, $S_{ab}$ takes the familiar form 
\begin{equation}\label{HadamardMatrix}
    S_{ab} =\frac{1}{\sqrt{2}}\begin{bmatrix}
        1 & 1\\
        1 & -1
    \end{bmatrix}.
\end{equation}

For a single qubit, where $d=2$, the operators $X$ and $Z$ generate the following actions on $\ket{a}$
\begin{equation}\label{XandZOperators}
\begin{split}
    X \ket{a} &= \ket{a+1\, (\!\!\!\!\!\!\mod 2)},\\
    Z \ket{a} &= \omega^a \ket{a},\\
\end{split}
\end{equation}
where the relevant root of unity is the one that appears in the eigenvalues of $Z$, namely $\omega \equiv \exp(2\pi i / d) = -1$ for $d=2$. We emphasize that this $\omega$ is the same as the one defined in Eq.\ \eqref{PauliQuditOperators}, and is distinct from the phase $\hat{\omega} \equiv \exp(2\pi i / D)$ introduced in the surrounding discussion. As defined, $\hat{\omega}$ helps define the global phase classes of the $n$-qudit Pauli group by the set $\{\hat\omega^r X^a Z^b\}$, and does not affect the action of $X$ or $Z$ on individual states $\ket{a}$. The operators $X$ and $Z$ are independent of level-rank $k$, and obey the relation
\begin{equation}
    XZ = \omega^{-1} ZX.
\end{equation}

As shown in Eqs.\ \eqref{PauliQuditStrings} and \eqref{PauliQuditCommutation}, the action of the single qudit Pauli group can be extended to arbitrary qudit number by constructing the tensor product $X^aZ^b$, up to some global phase. The operator $X^aZ^b$, and its phase multiples, given by 
\begin{equation}\label{QuditGroupSet}
   \{\omega^cX^aZ^b |c\in\mathbb{Z}_d\},
\end{equation}
define the $n$-qudit Pauli group. As shown by the relations in Eq.\ \eqref{PauliQuditCommutation}, the group in Eq.\ \eqref{QuditGroupSet} is isomorphic to the $2n$ commutative ring
\begin{equation}\label{CommRing}
   M_R = \mathbb{Z}_d \times \mathbb{Z}_d \times ... \times \mathbb{Z}_d,
\end{equation}
with multiplicative Pauli group action corresponding to multiplication in the ring. Accordingly, the $n$-qudit Pauli group corresponds to the disjoint union of $3$-manifolds. In this representation, the operator $Z$ is prepared~\cite{Salton:2016qpp} as a path-integral over the manifold where the boundary tori are identified with $\mathbb{T}^2$.

\subsection{Clifford Operators in $SU(2)_1$}\label{TopologicalClifford}

%Recall that, for dimension $d >2$, the Pauli matrices generalize to the operator form given by Eq.\ \eqref{PauliQuditOperators}. Following~\cite{Farinholt_2014}, for our case we have
%
%\begin{equation}
%    X = \sum_{x \in \mathbb{Z}_d}\ket{x+1}\bra{x}, \quad Z = \sum_{z \in \mathbb{Z}_d} \omega^z\ket{z}\bra{z}.
%\end{equation}
%
%When $D=2d$, since $d=2$, we get explicit results for the Clifford group.

Following our algebraic construction of the Pauli group in Section \ref{KacMoodyPauli}, we can likewise construct the qubit Clifford group as topological operators in the Kac-Moody algebra $SU(2)_1$. The qudit Clifford group is generated by the  Hadamard (QFT), phase, and $C_{Sum}$ gates, defined in Section \ref{QuditCliffordSection}. For $SU(2)_1$, the simplest non-trivial topological theory is the well-known Wess-Zumino-Witten (WZW) model for rational conformal field theories~\cite{Witten1984}. Constructing Clifford operators in the WZW model, we begin with the Hadamard gate $S_{ab}$ which is represented by a modular tensor category $S$-matrix. Interestingly, for $SU(2)_1$, this $S$-matrix is exactly the canonical Hadamard matrix
\begin{equation}\label{KacMoodyHadamard}
    S_{ab} = \frac{1}{\sqrt{2}}\begin{bmatrix} 1 & 1\\
    1 & -1 \\
    \end{bmatrix},
\end{equation}
as shown in Eq.\ \eqref{HadamardMatrix}.

The phase operator in $SU(2)_1$ can be built using the Kac-Moody modular transformation matrix $T_{ab}$, which we define below in terms of the central charge of the theory $c$, and conformal dimension $h_a$
\begin{equation}\label{ModularTransformationMatrix}
    T_{ab} = \exp \left[2\pi i \left(h_a - \frac{c}{24} \right) \right]\delta_{ab}.
\end{equation}
For $SU(2)_1$ the central charge $c=1$, and the conformal dimension $h_a$ is expressed as a function of the quadratic Casimir operator $C_2(a)$ as
\begin{equation}\label{ConformalDimension}
    h_a = \frac{C_2(a)}{3} = \frac{\frac{a}{4}(a+2)}{3}.
\end{equation}
We construct the phase operator $P_{ab}$ using the modular transform in Eq.\ \eqref{ModularTransformationMatrix} as
\begin{equation}
    P_{ab} = \left(\omega^{-1/2}\omega^{a(a+2)/6}T_{ab} \right),
\end{equation}
where again $\omega$ denotes the $d^{th}$ root of unity as in Eq.\ \eqref{GeneralizedPhase}.

For $d=2$, the $C_{Sum}$ operator from Eq.\ \eqref{CSum} performs the following action
\begin{equation}
    C_{Sum}\ket{a}\ket{b} = \ket{a+b, \mod 2}.
\end{equation}
We produce an algebraic representation for  $C_{Sum}$ by first conjugating the fusion tensor in Eq.\ \eqref{FusionMatrix} by Fourier transforms, generating a copy tensor. Contracting this copy tensor with the fusion tensor then builds the algebraic representation for $C_{Sum}$, as shown in Figure \ref{CSumGraph}. 

Importantly, as we describe above, the Fourier operator $S_{ab}$ and fusion tensor $N^c_{ab}$ are topologically constructible. Therefore for $SU(2)_1$, we can obtain any element of the Clifford group both topologically and algebraically. Stated directly, we can prepare any Clifford operator on $n$-qubits, for $SU(2)_1$, using a path-integral over some $3$-manifold $\mathcal{M}_3$ with $2n$ boundary tori. 

\section{$W_n$ States and Entropies in $SU(2)_1$}\label{WEntropies}

Previous work~\cite{Salton:2016qpp} has successfully constructed a  topological presentation for stabilizer states in $U(1)_k$, with $k$ odd, and their associated entanglement entropies. We extend this approach beyond stabilizer states to include more general classes of states, to include the state $W_n$, which are frequently prepared as initial states for quantum algorithms~\cite{Dicke1954,Stockton2003,Bastin2009,Munizzi:2023ihc}. We further explore the evolution of entanglement entropy under groups of quantum operators, such as the Clifford group, in a topological framework.

The $n$-qubit stabilizer states are $+1$ eigenstates of a maximally-invariant subgroup of the Pauli group, with order $2^n$. In quantum computing, stabilizer states are precisely the states reachable using only the Clifford group acting on any measurement basis state, e.g. $\ket{0}^{\otimes n}$. Clifford gates can be efficiently simulated on a classical computer in polynomial time, therefore computations using stabilizer states are inexpensive in quantum resource theory. In a $U(1)_k$ Chern-Simons theory, stabilizer states are characterized by their topology. Any stabilizer state can be prepared on a Hilbert space consisting of $n$ copies of a torus $\mathbb{T}^2$ using the topological representations of Clifford operators given in Section \ref{TopologicalClifford}. Any $n$-qubit Clifford operator $U \in \mathcal{C}_n$ can be prepared by a path integral over a $3$-dimensional manifold $\mathcal{M}_3$ with $2n$ boundary tori, where half the tori carry one orientation and the other half carry the opposite orientation~\cite{Salton:2016qpp}. The specific manifold used to produce $U$ can be acquired by gluing a solid torus $\mathcal{T}$ to each boundary copy of $\mathbb{T}^2$. Calculating stabilizer state entanglement entropy then follows as described in Section \ref{ChernSimonsEntropySection}.

Using a similar construction, we can also prepare states beyond the set of stabilizer states. As an example, consider the $n$-qubit $W$ state, i.e. the equal weight superposition over all single-site excitations, which is not a stabilizer state for $n \geq 3$. We prepare $W_n$ through the action of $X_i$ on $\ket{0}^{\otimes n}$ as
\begin{equation}\label{WStateConstruction}
    W_n = \frac{1}{\sqrt{n}} \sum_{i=1}^n X_i \ket{0}^{\otimes n},
\end{equation}
where $X_i \ket{j} = \ket{j+1, \mod 2}$ acts on the $i^{th}$ qubit as defined in Eq.\ \eqref{XandZOperators}. In $SU(2)_1$, the action of $X_i$ in Eq.\ \eqref{WStateConstruction} corresponds to the fusion matrix given by Eq.\ \eqref{FusionMatrix}, enabling a topological construction of $W_n$ in $SU(2)_1$. Moreover, we can extend this topological construction to all Dicke states $\ket{D_k^n}$, for which $W_n = \ket{D_1^n}$.

\subsection{A Topological Preparation for $W_n$ States}\label{TopologicalPreparationSection}

We now introduce a topological construction for $W_n$ states and their entanglement entropies in $SU(2)_1$ Chern-Simons theory. We begin by composing an operator generation for $\ket{W_n}$, which we then interpret in a topological framework. We then use the topological construction to compute the entanglement entropies of $W_n$.

As shown in Eq.\ \eqref{WStateConstruction}, we can construct $W_n$ through the action of $X_i$ on $\ket{0}^{\otimes n}$. The action of $X_j$ on a single-qudit basis state coincides, by Eq.\ \eqref{FusionMatrix}, with that of the matrix elements $N^{a+1}_{a,1}$ of the $SU(2)_1$ fusion tensor with one input fixed to the fundamental representation
\begin{equation}\label{XandFusion}
\begin{split}
        X_j \ket{a} &= \ket{a+1\, (\!\!\!\!\!\!\mod 2)},\\
        N^{a+1}_{a,1} \ket{a} &= \ket{a+1\, (\!\!\!\!\!\!\mod 2)}.
\end{split}
\end{equation}
The product $X_j X_j^\dagger = \mathbb{1}$ then equals $N^{a+1}_{a,1} (N^{a+1}_{a,1})^\dagger$ at the level of fusion tensors, which we use to give a topological realization of the diagonal entries of $\rho_{W_n}$. Constructing the density matrix $\rho_{W_n} \equiv \ket{W_n}\bra{W_n}$ by summing over the action of $X_j$ on the state $\ket{0}^{\otimes n}$ then gives
\begin{equation}\label{WDiag}
\begin{split}
    \rho_{W_n} &= \frac{1}{n}\sum_{j=1}^n X_j \ket{00...0}\bra{00...0} X_j^\dagger,\\
    &= \frac{1}{n}\sum_{j=1}^n \left(N_{j,1}^{j+1} \right) \ket{00...0}\bra{00...0} \left(N_{j,1}^{j+1} \right)^\dagger.
\end{split}
\end{equation}
As we demonstrate in Section \ref{WEntropySection}, the diagonal entries of Eq.\ \eqref{WDiag} are sufficient to topologically compute $W_n$ entropies~\cite{Schnitzer:2020tiv}.

\begin{figure}[h]
\centering
\includegraphics[width=7cm]{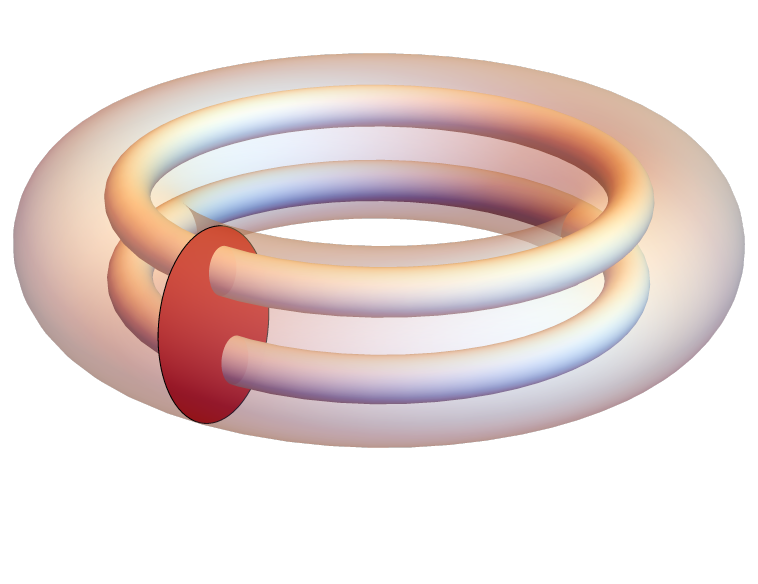}
\caption{Manifold $\eta$ consisting of a solid torus with two tori removed from its interior~\cite{Salton:2016qpp}. This manifold has three toroid boundaries. The associated fusion tensor $N_{j_1,j_2}^{j_3}$, and thereby $W_n$, can be prepared by path integration over $\eta$.}
\label{Manifold}
\end{figure}

We prepare $W_n$ topologically by constructing the fusion tensor $N_{j,1}^{j+1}$ from Eq.\ \eqref{WDiag} by path integration, similar to the protocol introduced in~\cite{Salton:2016qpp}. Consider the manifold $\eta$ in Figure \ref{Manifold}, composed of a solid torus with two tori removed from its interior.
The boundary of $\eta$ consists of three tori, which yields the corresponding fusion tensor $N_{j_1,j_2}^{j_3} = \delta_{j_3,j_1+j_2}$. 

Since $\eta$ is a manifold with boundary, path integration on $\eta$ corresponds to state preparation in the Hilbert space $\Hil = \partial\eta$ associated to the boundary field configuration~\cite{Salton:2016qpp}. When one of the solid tori in $\eta$ carries a Wilson loop in the fundamental (spin-1/2) representation of $SU(2)_1$, labeled $a = 1$ using the indexing in Section~\ref{KacMoodyPauli}, the path integral yields the fusion tensor with one index fixed to this value, and the fusion tensor simplifies to $N_{j_1, 1}^{j_1 + 1}$, which we write as $N_{j, 1}^{j+1}$ for simplicity. A graphical representation of $N_{j,1}^{j+1}$ is given by the three-legged tensor in Figure~\ref{TensorGraph}.
\begin{figure}
\begin{center}
\begin{tikzpicture}
    % Define the central point
    \coordinate (O) at (0,0);
    % Define the end points of the directed lines
    \coordinate (A) at (2,-1);
    \coordinate (B) at (-2,-1);
    \coordinate (C) at (0,2);
    % Draw the directed lines with centered arrows
    \draw[thick] (O) -- (A) node[midway, above] {};
    \draw[thick] (O) -- (B) node[midway, above] {};
    \draw[thick] (O) -- (C) node[midway, above] {};
    \node at (-4,.5) {$N_{j,1}^{j+1} = \delta_{j,j+1} = $};
    \node at (2.2,-1.2) {1};
    \node at (-2.2,-1.2) {j};
    \node at (0,2.2) {j+1};
\usetikzlibrary{calc}
    % Add arrows at the midpoint
    \draw[thick,->] (A) -- ($(O)!0.5!(A)$);
    \draw[thick,->] (B) -- ($(O)!0.5!(B)$);
    \draw[thick,->] (O) -- ($(O)!0.5!(C)$);
\end{tikzpicture}
\caption{Graphical representation of the fusion tensor $N_{j,1}^{j+1}$, which maps two Hilbert spaces into one. The tensor $N_{j,1}^{j+1}$ is prepared by path integration over the manifold $\eta$ in Figure \ref{Manifold}. This tensor encodes how maximally-entangled states, e.g. $W_n$, are prepared topologically using path integration.}
\label{TensorGraph}
\end{center}
\end{figure}

We likewise wish to prepare the adjoint operator $\left(N_{j,1}^{j+1}\right)^\dagger$, which corresponds to inverting the direction of each arrow in Figure \ref{TensorGraph}, generating a copy tensor. The action $\left(N_{j,1}^{j+1}\right) \cdot \left(N_{j,1}^{j+1}\right)^\dagger$ applied in Eq.\ \eqref{WDiag} is then obtained by gluing the fusion tensor from Figure \ref{TensorGraph} together with its copy~\cite{Salton:2016qpp}. The graphical representation for the resulting $\left(N_{j,1}^{j+1}\right) \cdot \left(N_{j,1}^{j+1}\right)^\dagger$ composition is given by Figure \ref{TorusAndGenus}.
\begin{figure}
\begin{center}
\begin{tikzpicture}[scale=2]
    % Draw the circle in arc segments with a break
    \draw[->, thick] (45:1) arc[start angle=45, end angle=90, radius=1];
    \draw[thick] (90:1) arc[start angle=90, end angle=180, radius=1];
    \draw[thick] (-45:1) arc[start angle=-45, end angle=-90, radius=1];
    \draw[<-,thick] (-90:1) arc[start angle=-90, end angle=-180, radius=1];

    % Rhombus inside
    \coordinate (A) at (45:1);
    \coordinate (B) at (1.4, 0);
    \coordinate (C) at (0, 0);
    \coordinate (D) at (315:1);
  
    \draw[thick] (B) -- (A) node[midway, above] {};
    \draw[thick] (C) -- (A) node[midway, above] {};
    \draw[thick] (D) -- (C) node[midway, above] {};
    \draw[thick] (D) -- (B) node[midway, above] {};
    
    \node at (-2,0) {$\left(N_{j,1}^{j+1}\right) \cdot \left(N_{j,1}^{j+1}\right)^\dagger$ =};
    \node at (1.55,0) {$1$};
    \node at (-.15,0) {$j$};
    \node at (0,1.2) {$j+1$};
    \node at (0,-1.2) {$j+1$};
    
  \draw[thick, postaction={decorate}, decoration={markings, mark=at position 0.5 with {\arrow{Latex}}}] (B) -- (A);
  \draw[thick, postaction={decorate}, decoration={markings, mark=at position 0.5 with {\arrow{Latex}}}] (C) -- (A);
  \draw[thick, postaction={decorate}, decoration={markings, mark=at position 0.5 with {\arrow{Latex}}}] (D) -- (C);
  \draw[thick, postaction={decorate}, decoration={markings, mark=at position 0.5 with {\arrow{Latex}}}] (D) -- (B);
\end{tikzpicture}
\caption{Graphical representation for the action $\left(N_{j,1}^{j+1}\right) \cdot\left(N_{j,1}^{j+1}\right)^\dagger$, as in Eq.\ \eqref{WDiag}, composed of $N_{j,1}^{j+1}$ with its copy tensor $\left(N_{j,1}^{j+1}\right)^\dagger$ glued together. Both fusion tensors are prepared using path integration along $\eta$, enabling a topological construction for $W_n$.}
\label{TorusAndGenus}
\end{center}
\end{figure}
Summing over $\left(N_{j,1}^{j+1}\right) \cdot\left(N_{j,1}^{j+1}\right)^\dagger$, for the number of single-site excitations, generates $W_n$. Each term in the sum in Eq.\ \eqref{WDiag} corresponds to a single copy of Figure \ref{TorusAndGenus}.

The action of $N_{j,1}^{j+1}$ on a state is defined in Eq.\ \eqref{FusionMatrix}, and topologically obtained by path integration on the manifold $\eta$. Computing Eq.\ \ref{WDiag} topologically we first recall that $N_{j,1}^{j+1} = \delta_{j,j+1}$, therefore
\begin{equation}\label{SumXj}
    \frac{1}{n}\sum_{j=1}^n X_j \cdot X_j^\dagger = \frac{1}{n}\sum_{j=1}^n \left(N_{j,1}^{j+1} \right) \cdot \left(N_{j,1}^{j+1} \right)^\dagger = \frac{n}{n} = 1.
\end{equation}
The sum in Eq.\ \eqref{SumXj} is constructed on the union of $n$ tori, which itself is a torus of genus $n$. This $n$-genus torus is represented by $n$ copies of the graphical representation in Figure \ref{TorusAndGenus}. Computing Eq.\ \eqref{SumXj} topologically then corresponds to a path integral over $n$ copies of $\eta$, giving%
\footnote{In general, the sum in Eq.\ \eqref{PathIntegralUnion} gives $Z\left(-\eta \cup_{\partial \eta} \eta \right) = k^2$, where $k$ is the level of the Chern-Simons theory. In this paper we consider theories for which $k=1$.} %
\begin{equation}\label{PathIntegralUnion}
    \left(N_{j,1}^{j+1} \right) \left(N_{j,1}^{j+1} \right)^\dagger = Z\left(-\eta \cup_{\partial \eta} \eta \right).
\end{equation}
Eq.\ \eqref{PathIntegralUnion} expresses the contraction $\left(N_{j,1}^{j+1} \right) \left(N_{j,1}^{j+1} \right)^\dagger$ as a single partition function over the closed manifold $-\eta \cup_{\partial \eta} \eta$. We emphasize that this identity is a topological re-expression of a single scalar quantity, not the single-replica entropy formula of Eq.\ \eqref{TopologicalEE}. As discussed in Section~\ref{ChernSimonsEntropySection}, Eq.\ \eqref{TopologicalEE} returns the second R\'enyi entropy in general and equals the von Neumann entropy only when the entanglement spectrum is flat. The $W_n$ states considered here are not stabilizer states for $n \geq 3$ and do not have flat spectrum; we therefore compute their von Neumann entropies in Section~\ref{WEntropySection} directly from the reduced density matrix and use Eq.~\eqref{PathIntegralUnion} only to recast the resulting closed-form scalar in topological language.

In this section, we describe the topological preparation of $W_n$ states. We utilize the interpretation of the generalized $X$ operator in Eq.\ \eqref{PauliQuditOperators} as the fusion tensor defined in Eq.\ \eqref{XandFusion}. This fusion tensor can be realized by a path integral over a solid torus with three boundary tori, illustrated in Figure \ref{Manifold}. When the manifold includes a Wilson loop operator through one of its tori, the fusion tensor simplifies to the form shown in Figure \ref{TensorGraph}. The contraction of $N_{j,1}^{j+1}$ with its adjoint admits a path-integral realization over the closed manifold $-\eta \cup_{\partial\eta} \eta$, providing the fundamental topological object that we use to compute the entanglement entropy of $W_n$ in closed form.

\subsection{Topological Calculation of $W_n$ Entropies}\label{WEntropySection}

We now use the construction detailed in Section \ref{TopologicalPreparationSection} to provide a topological calculation for $W_n$ entanglement entropies. We compute the entanglement entropy as the von Neumann entropy of the reduced density matrix corresponding to a chosen subsystem of $W_n$, evaluated directly from its closed-form spectrum. The path integral provides a topological re-expression of the resulting entropy in terms of partition functions on copies of the manifold $\eta$. We demonstrate how the elements of the reduced density matrix are generated by copies of $\eta$, and how the von Neumann entropy then admits a manifestly topological form.

For the multipartite state $W_n$, and a chosen subsystem $A \in W_n$, entanglement entropy of party $A$ is computed using the von Neumann entropy $S_A$ defined in Eq.\ \eqref{SubsystemEntropyDefinition}. Moreover, when subregion $A$ corresponds to a single qubit, the reduced density matrix $\rho_A$ has the form
\begin{equation}\label{WReducedDensity}
    \rho_A = \frac{1}{n}\ket{11...1}\bra{11...1} + \frac{n-1}{n}\ket{00...0}\bra{00...0}. 
\end{equation}
As a result, computing $S_A$ from Eq.\ \eqref{WReducedDensity} admits the simplified expression~\cite{Schnitzer:2022exe, Munizzi:2023ihc}
\begin{equation}\label{WnEntropies}
    S_A = -\frac{1}{n}\log_d \left(\frac{1}{n} \right) - \frac{n-1}{n} \log_d \left(\frac{n-1}{n} \right).
\end{equation}

We compute the entanglement entropy $S_A$ from the density matrix%
\footnote{For manifold $\eta$, the reduced density matrix may be written $\rho_A = \tr \partial_{\mathcal{\eta}}\backslash A$.} %
$\rho_A$. The elements of $\rho_A$ are populated directly by the action of $X_i$ on $\ket{0}^{\otimes n}$, as in Eq.\ \eqref{WStateConstruction}, allowing Eq.\ \eqref{WReducedDensity} to be expressed as
\begin{equation}\label{OperatorReducedDensity}
    \rho_A = \frac{1}{n}\left(X_A \cdot X_A^\dagger + (n-1)\mathbb{1} \right) \ket{00...0}\bra{00...0}.
\end{equation}
Since $X_A$ can be expressed topologically, the entropies $S_A$ likewise admit a topological interpretation. Calculating $S_A$ using Eq.\ \eqref{WnEntropies}, and employing the operator construction in Eq.\ \eqref{OperatorReducedDensity}, we can express $W_n$ entanglement entropies in terms of $X_A$
\begin{equation}\label{XAentropy}
    S_A = \tr \left[\frac{X_A \cdot X_A^\dagger}{n} \log_d \frac{X_A \cdot X_A^\dagger}{n} \right] + \tr \left[\frac{n-1}{n}\mathbb{1} \log_d \left(\frac{n-1}{n}\mathbb{1} \right) \right]. 
\end{equation}
The construction in Eq.\ \eqref{XAentropy}, upon interpreting $X_A$ as a fusion tensor, enables a topological interpretation of entanglement entropy for states $W_n$. 

We now use the relationship between $X_A$ and fusion tensor $N_{j,1}^{j+1}$ to compose a topological representation for $S_A$. Combining Eq.\ \eqref{PathIntegralUnion} with Figure \ref{TorusAndGenus}, we express the path integral $ Z\left(\eta \cup_{\partial\eta}\eta \right)$ using the graphical representation in Figure \ref{GraphRepOfPathInt}.
\begin{figure}
\begin{center}
\begin{tikzpicture}[scale=2]
    % Draw the circle in arc segments with a break
    \draw[->, thick] (45:1) arc[start angle=45, end angle=90, radius=1];
    \draw[thick] (90:1) arc[start angle=90, end angle=180, radius=1];
    \draw[thick] (-45:1) arc[start angle=-45, end angle=-90, radius=1];
    \draw[<-,thick] (-90:1) arc[start angle=-90, end angle=-180, radius=1];

    % Rhombus inside
    \coordinate (A) at (45:1);
    \coordinate (B) at (1.4, 0);
    \coordinate (C) at (0, 0);
    \coordinate (D) at (315:1);
  
    \draw[thick] (B) -- (A) node[midway, above] {};
    \draw[thick] (C) -- (A) node[midway, above] {};
    \draw[thick] (D) -- (C) node[midway, above] {};
    \draw[thick] (D) -- (B) node[midway, above] {};

    \node at (-2.2,0) {$\left(N_{j,1}^{j+1}\right) \cdot \left(N_{j,1}^{j+1}\right)^\dagger$ =};
    \node at (2.5,0) {$= Z\left(-\eta \cup_{\partial\eta}\eta \right)$};
    \node at (1.15,.6) {$1$};
    \node at (1.15,-.6) {$1$};
    \node at (0.15,.6) {$j$};
    \node at (0.15,-.6) {$j$};
    \node at (0,1.2) {$j+1$};
    \node at (0,-1.2) {$j+1$};
    
  \draw[thick, postaction={decorate}, decoration={markings, mark=at position 0.5 with {\arrow{Latex}}}] (B) -- (A);
  \draw[thick, postaction={decorate}, decoration={markings, mark=at position 0.5 with {\arrow{Latex}}}] (C) -- (A);
  \draw[thick, postaction={decorate}, decoration={markings, mark=at position 0.5 with {\arrow{Latex}}}] (D) -- (C);
  \draw[thick, postaction={decorate}, decoration={markings, mark=at position 0.5 with {\arrow{Latex}}}] (D) -- (B);
\end{tikzpicture}
\caption{The action of $\left(N_{j,1}^{j+1}\right) \cdot \left(N_{j,1}^{j+1}\right)^\dagger$ is topologically prepared as a path integral over the glued manifold $-\eta \cup_{\partial\eta}\eta$. The final manifold is formed by joining two copies of $\eta$, depicted in Figures \ref{Manifold} and \ref{TorusAndGenus}, along their common boundary with opposite orientation. This construction gives a topological interpretation for the operator used to construct $W_n$.}
\label{GraphRepOfPathInt}
\end{center}
\end{figure}

Taking the sum over multiple attached copies of Figure \ref{GraphRepOfPathInt}, as in Eq.\ \eqref{SumXj}, corresponds to summing over path integrals performed on each copy of $-\eta \cup \eta$, as by Eq.\ \eqref{PathIntegralUnion}. Figure \ref{SumGraphRepOfPathInt} gives a graphical depiction of this sum over replicas of $-\eta \cup \eta$.
\begin{figure}
\begin{center}
\begin{tikzpicture}[scale=2]
    % Draw the circle in arc segments with a break
    \draw[->, thick] (45:1) arc[start angle=45, end angle=90, radius=1];
    \draw[thick] (90:1) arc[start angle=90, end angle=180, radius=1];
    \draw[thick] (-45:1) arc[start angle=-45, end angle=-90, radius=1];
    \draw[<-,thick] (-90:1) arc[start angle=-90, end angle=-180, radius=1];

    % Rhombus inside
    \coordinate (A) at (45:1);
    \coordinate (B) at (1.4, 0);
    \coordinate (C) at (0, 0);
    \coordinate (D) at (315:1);
  
    \draw[thick] (B) -- (A) node[midway, above] {};
    \draw[thick] (C) -- (A) node[midway, above] {};
    \draw[thick] (D) -- (C) node[midway, above] {};
    \draw[thick] (D) -- (B) node[midway, above] {};
    
    \node at (-1.5,0) {$\sum_{j=1}^n$};
    \node at (3,0) {$=\sum_{j=1}^n Z\left(-\eta_j \cup_{\partial_{\eta_j}} \eta_j \right)$};
    \node at (1.15,.6) {$1$};
    \node at (1.15,-.6) {$1$};
    \node at (0.15,.6) {$j$};
    \node at (0.15,-.6) {$j$};
    \node at (0,1.2) {$j+1$};
    \node at (0,-1.2) {$j+1$};
    
  \draw[thick, postaction={decorate}, decoration={markings, mark=at position 0.5 with {\arrow{Latex}}}] (B) -- (A);
  \draw[thick, postaction={decorate}, decoration={markings, mark=at position 0.5 with {\arrow{Latex}}}] (C) -- (A);
  \draw[thick, postaction={decorate}, decoration={markings, mark=at position 0.5 with {\arrow{Latex}}}] (D) -- (C);
  \draw[thick, postaction={decorate}, decoration={markings, mark=at position 0.5 with {\arrow{Latex}}}] (D) -- (B);
\end{tikzpicture}
\caption{Evaluating the sum in Eq.\ \eqref{SumXj} corresponds to summing over copies of $-\eta \cup_{\partial_{\eta}} \eta$, each attached at their shared boundaries. Likewise, summing over the results of independent path integrals on each $-\eta \cup_{\partial_{\eta}} \eta$ gives a topological construction for Eq.\ \eqref{SumXj}.}
\label{SumGraphRepOfPathInt}
\end{center}
\end{figure}
%%and thereby $-\eta \cup \eta$ is a manifold with $3 \times 3$ torus boundaries. 
Each copy of $\eta$ has $3$ torus boundaries, and the union of a single $(-\eta \cup_{\partial_n} \eta )$ has genus $1$, which comes from the union of boundaries in Figure \ref{Manifold}. Therefore, computing the sum in Figure \ref{SumGraphRepOfPathInt} above we have
\begin{equation}\label{PathIntFusionTensor}
    \sum_{j=1}^n Z(-\eta_j \cup_{\partial_{n_j}} \eta_j) = n,
\end{equation}
where each summand is disjoint. This construction differs from a similar construction presented in Figure 4(b) of~\cite{Salton:2016qpp}.

Expanding the $\log$ in Eq.\ \eqref{WnEntropies} and simplifying, we can rewrite the single-region entropies $S_A$ for $W_n$ as
\begin{equation}\label{RewrittenWnEntropy}
    S_A = \log_d(n) - \frac{n-1}{n}\log_d(n-1).
\end{equation}
Using the results of our path integration over the manifold in Figure \ref{SumGraphRepOfPathInt}, we arrive at a topological realization of Eq.\ \eqref{RewrittenWnEntropy}. Combining Eq.\ \eqref{PathIntFusionTensor} with Eq.\ \eqref{RewrittenWnEntropy} we have
\begin{equation}\label{TopologicalSA}
    S_A = \log_d\left[\sum_{j=1}^n Z_j\left(-\eta_j \cup_{\partial \eta_j} \eta_j\right) \right] - \frac{1}{Z\left(-\eta \cup_{\partial \eta} \eta\right)}\log_d\left[\sum_{j=1}^{n-1} Z_j\left(-\eta_j \cup_{\partial \eta_j} \eta_j\right)\right].
\end{equation}
Eq.\ \eqref{TopologicalSA} gives a topological realization of entanglement entropies for single parties in the state $W_n$.

For multiparty subsystems of $W_n$, the entanglement entropy has a form similar to that in Eq.\ \eqref{WnEntropies}. For the $n$-partite state $W_n$, the entanglement entropy $S_\ell$ of any $\ell$-partite subsystem is given by
\begin{equation}\label{GeneralWnEntropies}
    S_\ell = -\frac{\ell}{n}\log_d \left(\frac{\ell}{n} \right) - \frac{n-\ell}{n} \log_d \left(\frac{n-\ell}{n} \right).
\end{equation}
Therefore, in the case when $\ell = AB$ we have
\begin{equation}
\begin{split}
    S_{AB} &= \frac{2}{\etaUnionSum} \log_d \left[\frac{\etaUnionSum}{2}
     \right]\\
    \quad &+ \frac{1}{\sum_{j=1}^2 Z(\eta_j \cup_{\partial_{\eta_j}} \eta_j)}\log_d \left[\sum_{j=1}^2 Z(\eta_j \cup_{\partial_{\eta_j}} \eta_j) \right],
\end{split}
\end{equation}
and similarly for $\ell = ABC$,
\begin{equation}
\begin{split}
    S_{ABC} &= \frac{3}{\etaUnionSum} \log_d \left[\frac{\etaUnionSum}{3}
     \right]\\
    \quad &+ \frac{1}{\sum_{j=1}^3 Z(\eta_j \cup_{\partial_{\eta_j}} \eta_j)}\log_d \left[\sum_{j=1}^3 Z(\eta_j \cup_{\partial_{\eta_j}} \eta_j \right].
\end{split}
\end{equation}
Thus for a general $\ell$-party subsystem of $W_n$ we have the following topological representation for $S_\ell$
\begin{equation}
\begin{split}
    S_{\ell} &= \frac{\ell}{\etaUnionSum} \log_d \left[\frac{\etaUnionSum}{\ell}
     \right]\\
    \quad &+ \frac{1}{\sum_{j=1}^\ell Z(\eta_j \cup_{\partial_{\eta_j}} \eta_j)}\log_d \left[\sum_{j=1}^\ell Z(\eta_j \cup_{\partial_{\eta_j}} \eta_j \right].
\end{split}
\end{equation}
We now compute an explicit example for the bipartite state $W_2$.

\paragraph{Example}

The simplest example we can consider is a topological representation for the $2$-qubit maximally-entangled state $\ket{W_2}$, and its associated entanglement entropies. The state $\ket{W_2} = 1/\sqrt{2}\left(\ket{01}+\ket{10}\right)$ is a stabilizer state, and therefore can be prepared in the $2$-torus Hilbert space $\mathcal{H}^{\otimes 2}_{\mathbb{T}^2}$, using a single path integral over the manifold $\eta$ in Figure \ref{Manifold}. Using 
Eq.\ \eqref{XandZOperators}, we construct $\ket{W_2}$ as
\begin{equation}
    W_2 = \frac{1}{\sqrt{2}} \left(X_1 \otimes \mathbb{1}_2 + \mathbb{1}_1 \otimes X_2\right)\ket{00}.
\end{equation}
Each operator $X_i$ is produced by a path integral over a single copy of $\eta$. The components of $\ket{W}_2$ are computed via the expectation value of Wilson loop operators in $\mathbb{S}^2 \times \mathbb{S}^1$, each inserted along a curve $C_i$ passing through the core of each torus. This expectation value is proportional to the fusion tensor $N_{j,1}^{j + 1}$, and thereby
\begin{equation}\label{WilsonLoop}
    \langle \mathcal{W}(C_1,R_{j})\mathcal{W}(C_2,R_{1})\mathcal{W}(C_3,R^*_{j+1}) \rangle_{\mathbb{S}^2 \times \mathbb{S}^1} \propto N_{j,1}^{j + 1},
\end{equation}
where the $*$ superscript denotes the representation change acquired when changing orientation and passing to the dual Hilbert space $\mathcal{H}^*_{\mathbb{T^2}} \cong \mathcal{H}_{-\mathbb{T^2}}$. Therefore we have an expression for $\ket{W_2}$ as 
\begin{equation}
    \ket{W_2} = \frac{1}{\sqrt{2}} \left(\mathcal{W}(C_1,a_1)\mathbb{1}_2 + \mathbb{1}_1\mathcal{W}(C_2,a_2) \right)_{S^2 \times S^1},
\end{equation}
where the action of the identity operator is equivalent to
\begin{equation}
    \mathbb{1}_i = \mathcal{W}(C_i,a_i = 0) = \delta_{0,0}.
\end{equation}
Since $\ket{W_2}$ is a stabilizer state, it can be prepared as $\ket{W_2} = C\ket{0}^{\otimes 2}$, where $C$ is a Clifford operator. The corresponding operator $C$ can be prepared by a path integral over a genus-$2$ manifold%
\footnote{Note that the state $\ket{W_2}$ is special in that a single manifold of genus-$2$ is sufficient to topologically prepare the state. For $n\geq3$ the state $\ket{W_n}$ is no longer a stabilizer state, and further topological constructions require higher genus surfaces with multiple copies.} %
$\mathcal{M}$, composed of two tori glued together. Computing entanglement entropy $S_A$ for $\ket{W_2}$, the expression in Eq.\ \eqref{TopologicalSA} reduces to 
\begin{equation}
\begin{split}
    S_A &= \log_2\left[\sum_{j=1}^2 Z_j\left(-\eta_j \cup_{\partial \eta_j} \eta_j\right) \right] - \frac{1}{Z\left(-\eta \cup_{\partial \eta} \eta\right)}\log_2\left[Z\left(-\eta \cup_{\partial \eta} \eta \right)\right],\\
    & = \log_2[2] - \log_2[1],\\
    &=1.
\end{split}
\end{equation}

In this section we described the construction of $W_n$ from the all-zero state using the generalized $X$ operator, which we interpret as a fusion matrix in $SU(2)_1$. We demonstrated how this fusion tensor can be topologically prepared by a path integral over a manifold consisting of a single solid torus with two tori removed from its interior. Topologically building the fusion tensor, along with its associated copy tensor, enabled a path integral representation for $W_n$ construction. Furthermore, since we compute entanglement entropy from the reduced density matrix of $W_n$, we likewise provided a similar topological construction for $W_n$ entropies and computed a specific example for the state $W_2$. In the next section we describe a topological representation for the Clifford group, which we use to describe Clifford orbits of $W_n$.

\section{Clifford Orbits in $SU(2)_1$}\label{CliffordKacMoody}

In this section we give a topological presentation for Clifford group action in $SU(2)_1$. We begin by first defining algebraic constructions for each Clifford generator, as well as their associated graphical representations. We provide a topological construction for our algebraic Clifford operators, which we then use to describe the Clifford action on states $W_n$.

\subsection{An Algebraic Representation of Generalized Clifford Operators}

Recall that the single-qubit Clifford group $\mathcal{C}_1$ is generated by the Hadamard and phase operations, with matrix representations given by Eq.\ \eqref{CliffordGates}. In $SU(2)_1$, the generalized Hadamard gate $H$ is the modular transformation matrix for $SU(2)_1$, introduced in Eq.\ \eqref{HadamardMatrix}, given by
\begin{equation}
    H = S_{ab} =\frac{1}{\sqrt{2}}\begin{bmatrix}
        1 & 1\\
        1 & -1
    \end{bmatrix},
\end{equation}
in the standard normalization of $a=0, b=1$. Recall also from Eq.\ \eqref{GeneralizedPhase}, with even $d$, we can write the generalized phase gate $\overline{P}$ as
\begin{equation}
    \overline{P} = \omega^{\frac{j^2}{2}} \ket{j} = \begin{bmatrix}
        1 & 0\\
        0 & \omega^{1/2}
    \end{bmatrix},
\end{equation}
where again $\omega = \exp(i\pi) = -1$. Extending to $n>1$, the $C_{sum}$ operator from Eq.\ \eqref{CSum} is required to generate the group $\mathcal{C}_n$. Recall $C_{sum}$ is defined by the following action 
\begin{equation}
    C_{sum} \ket{i}\ket{j} = \ket{i+j; \mod 2}.
\end{equation}

We construct an algebraic representation~\cite{Salton:2016qpp} for $C_{sum}$ by first building the copy tensor, conjugating the adjoint of the fusion tensor $N$ with the modular transformation matrix $S_{ab}$. Contracting this copy tensor with the fusion tensor builds the graphical representation for $C_{sum}$ shown below. 
\begin{figure}[h]
\begin{center}
\begin{tikzpicture}[thick, scale=1]

  % Left vertical wire, split around S† and S boxes
  \draw (0,0) -- (0,0.5);
  \draw (0,1.1) -- (0,2);
  \draw (0,2) -- (0,2.9);
  \draw (0,3.5) -- (0,4);

  % Right vertical wire (no boxes)
  \draw (3,0) -- (3,4);

  % Horizontal wire, split around S box
  \draw[->] (0,2) -- (1.3,2);
  \draw[->] (1.9,2) -- (3,2);

  % S† gate at bottom left
  \draw (-0.4,0.5) rectangle (0.4,1.1);
  \node at (0,0.8) {\( S^\dagger \)};

  % S gate at top left
  \draw (-0.4,2.9) rectangle (0.4,3.5);
  \node at (0,3.2) {\( S \)};

  % S gate on horizontal line
  \draw (1.3,1.6) rectangle (1.9,2.4);
  \node at (1.6,2) {\( S \)};

  % Dots at intersection points
  \filldraw (0,2) circle (2pt);
  \filldraw (3,2) circle (2pt);

  % Arrows for inputs
  \draw[->] (0,-0.5) -- (0,0);  \node at (0,-0.7) {\( j \)};
  \draw[->] (3,-0.5) -- (3,0);  \node at (3,-0.7) {\( i \)};

  % Arrows for outputs
  \draw[->] (0,4) -- (0,4.5);  \node at (0,4.7) {\( j \)};
  \draw[->] (3,4) -- (3,4.5);  \node at (3,4.7) {\( i + j \)};
  \node at (.75,1.7) {\( j \)};

  % Labels on vertical lines
  \node at (-0.3,2.1) {\( N \)};
  \node at (3.3,2.1) {\( N \)};
  \end{tikzpicture}
  \caption{Graphical representation of $C_{sum}$ operator, built by contracting the fusion tensor $N$ with its copy tensor. The modular transformation $S$ is applied along each outgoing leg of $N$, with $S^\dagger$ applied along the single incoming leg.}
\label{CSumGraph}
\end{center}
\end{figure}

For even dimension $d$, we need to introduce an overall phase $\omega^{1/2(i+j)}$. Following Eq.\ \eqref{CSum}, for $d=2$, the gate $C_{sum}$ then becomes
\begin{equation}\label{CSumEven}
    C_{sum} \ket{i}\ket{j} = \omega^{1/2(i+j)} \ket{i}\ket{i+j; \mod 2}, 
\end{equation}
which yields an adjustment to Figure \ref{CSumGraph} to account for the action of $\overline{P}$. Figure \ref{CSumPhase} depicts the graphical construction of $C_{sum}$ for $d=2$.
\begin{figure}[h]
\begin{center}
\begin{tikzpicture}[thick, scale=1]

  % Left vertical wire (split around S† and S)
  \draw (0,0) -- (0,0.5);
  \draw (0,1.1) -- (0,2);
  \draw (0,2) -- (0,2.9);
  \draw (0,3.5) -- (0,4);

  % Horizontal wire with arrow
  \draw[->] (0,2) -- (1.3,2);
  \draw[->] (1.9,2) -- (3,2);

  % Boxes on left wire
  \draw (-0.4,0.5) rectangle (0.4,1.1);
  \node at (0,0.8) {\( S^\dagger \)};

  \draw (-0.4,2.9) rectangle (0.4,3.5);
  \node at (0,3.2) {\( S \)};

  % Middle S gate
  \draw (1.3,1.6) rectangle (1.9,2.4);
  \node at (1.6,2) {\( S \)};

  % P gate (on right vertical wire)
  \draw (3,3.2) circle (0.3);
  \node at (3,3.2) {\( P \)};  
  
  % Right vertical wire (split around P gate)
  \draw (3,0) -- (3,2.9);
  \draw (3,3.5) -- (3,4);

  % Dots at junctions
  \filldraw (0,2) circle (2pt);
  \filldraw (3,2) circle (2pt);

  % Input arrows
  \draw[->] (0,-0.5) -- (0,0); \node at (0,-0.7) {\( j \)};
  \draw[->] (3,-0.5) -- (3,0); \node at (3,-0.7) {\( i \)};

  % Output arrows
  \draw[->] (0,4) -- (0,4.5); \node at (0,4.7) {\( j \)};
  \draw[->] (3,4) -- (3,4.5); \node at (3.1,4.7) {\( i + j \)};

  % N labels
  \node at (-0.3,2.1) {\( N \)};
  \node at (3.3,2.1) {\( N \)};

  % Horizontal label
  \node at (.75,1.7) {\( j \)};

  % Omega phase label further to the right
  \node at (4.2,3.1) {\( \omega^{\frac{1}{2}(i + j)} \)};

\end{tikzpicture}
\caption{Graphical representation of $C_{sum}$ operator for dimension $d$ even. Following Eq.\ \eqref{CSumEven}, when $d$ is even an additional phase gate is applied to the outgoing leg of the contracted tensors.}
\label{CSumPhase}
\end{center}
\end{figure}
where a phase gate is applied to the output leg of the contracted tensors.

\subsection{A Topological Realization of Clifford Orbits}

Using the algebraic construction of Clifford operators introduced in Figures~\ref{CSumGraph}--\ref{CSumPhase}, we now develop a topological representation for Clifford group action on $W_n$ states. We begin by expressing the action of generalized Clifford operators as sums over manifolds, such as Figure \ref{Manifold}. This presentation allows for a translation between the canonical operator presentation for Clifford group action, into a path-integral representation over $3$-manifolds with Wilson loop insertions.

We begin by recalling the action of the quantum Fourier transform $S$, defined in Eq.\ \eqref{QFT}, which transforms a state $\ket{a}$ as
\begin{equation}
    \begin{split}
        a=0: S\ket{0} &= \frac{1}{\sqrt{2}} \sum_{b=0}^1 \ket{b} = \frac{1}{\sqrt{2}}\left(\ket{0} + \ket{1} \right),\\
        a=1: S\ket{1} &= \frac{1}{\sqrt{2}} \sum_{b=0}^1 \omega^b\ket{b} = \frac{1}{\sqrt{2}}\left(\ket{0} - \ket{1} \right).
    \end{split}
\end{equation}
The operator $S$ realizes the action of the Hadamard gate, interchanging $X$ and $Z$ up to a phase. Explicitly, we have the relation $SX \ket{a} = ZS\ket{a}$, which enables the actions of $SX \ket{0}$ and $SX \ket{1}$ to be written as
\begin{equation}\label{BasisPresent}
\begin{split}
    SX \ket{0} &= ZS\ket{0},\\
    &= \frac{1}{\sqrt{2}} Z\ket{0} + \frac{1}{\sqrt{2}} Z\ket{1},\\
    &= \frac{1}{\sqrt{2}}\left[\ket{0} - \ket{1}\right],\\
        \quad \\
    SX \ket{1} &= ZS\ket{1},\\
    &= \frac{1}{\sqrt{2}} Z\ket{0} - \frac{1}{\sqrt{2}} Z\ket{1},\\
    &= \frac{1}{\sqrt{2}}\left[\ket{0} + \ket{1}\right].
\end{split}
\end{equation}

\begin{figure}[h]
    \begin{center}
        \begin{overpic}[width=7cm]{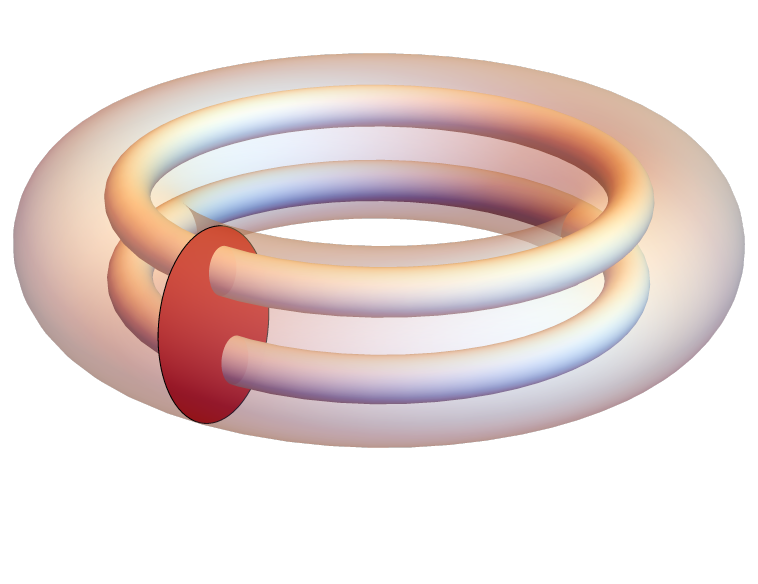}
            \put (31,38) {1}
            \put (32,27) {$a$}
            \put (102,40) {$= N_{a,1}^{a+1}$}
        \end{overpic}
    \caption{Manifold realizing the fusion tensor $= N_{a,1}^{a+1}$. Wilson loop expectation values computed on this manifold reproduce the algebraic relations for Clifford operators.}
    \label{WnManifold}
    \end{center}
\end{figure}

Topologically, the action of $X$ is implemented using the fusion kernel $N_{a,1}^{a+1}$, illustrated in Figure~\ref{WnManifold}. In this representation, the insertion of Wilson loop operators along curves passing through the cores of the interior tori encode this operator action on the qubit states in accordance with Eq.\ \eqref{WilsonLoop}.

The state $\ket{a} \in \mathcal{H}_{\mathbb{T}^2}$ associated to the integrable representation $a$ is prepared, in the topological language of Section~\ref{TopologicalPreparationSection}, by path integration over the manifold of Figure~\ref{WnManifold} with a Wilson loop along the core in the representation $a$~\cite{Salton:2016qpp}. In this notation, the action of an operator $\mathcal{O}$ on $\mathcal{H}_{\mathbb{T}^2}$ corresponds to a linear combination of such path integrals, with coefficients given by the matrix elements of $\mathcal{O}$ in the integrable basis. We use the convention where $\mathcal{W}(C, R)$ denotes a Wilson loop operator insertion in the representation $R$, while $\ket{a}$ denotes the corresponding boundary state, related by the path integral. Linearity of the path integral in the operator inserted then gives, for any linear combination of representations,
\begin{equation}\label{WnWilsonLoop}
    \Big\lvert \sum_{a} c_a\, R_a \Big\rangle \;=\; \sum_{a} c_a\, \ket{R_a},
\end{equation}
where $c_a$ are complex coefficients and the sum on the left-hand side is to be understood as a formal linear combination of operator insertions, not a single representation.

\begin{figure}[h]
  \begin{minipage}{.2\textwidth}
    \large{$SX\ket{0}= \frac{1}{\sqrt{2}}$}
  \end{minipage}
  \begin{minipage}{.35\textwidth}
        \begin{overpic}[width=5cm]{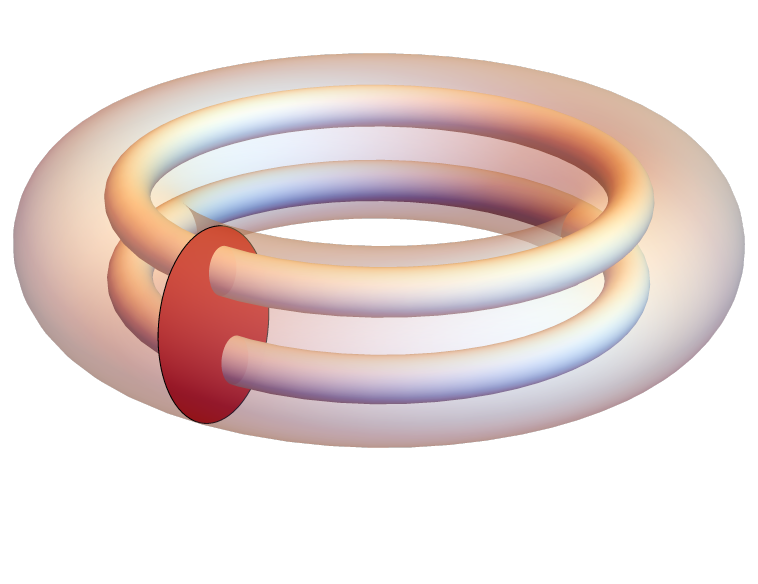}
            \put (31,38) {1}
            \put (32,27) {0}
            \put (-10,40) {\Huge{$\{$}}
        \end{overpic}
  \end{minipage}
  \begin{minipage}{.05\textwidth}
    \qquad \Large{$-$}
  \end{minipage}
    \begin{minipage}{.35\textwidth}
        \begin{overpic}[width=5cm]{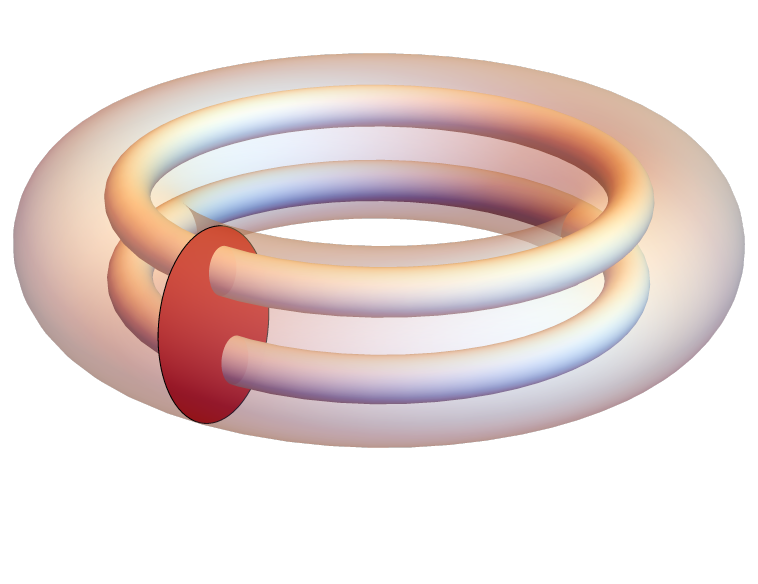}
            \put (31,38) {1}
            \put (32,27) {1}
            \put (100,40) {\Huge{$\}$}}
        \end{overpic}
  \end{minipage}\\
  \begin{minipage}{.2\textwidth}
    \large{$SX\ket{1}= \frac{1}{\sqrt{2}}$}
  \end{minipage}
  \begin{minipage}{.35\textwidth}
        \begin{overpic}[width=5cm]{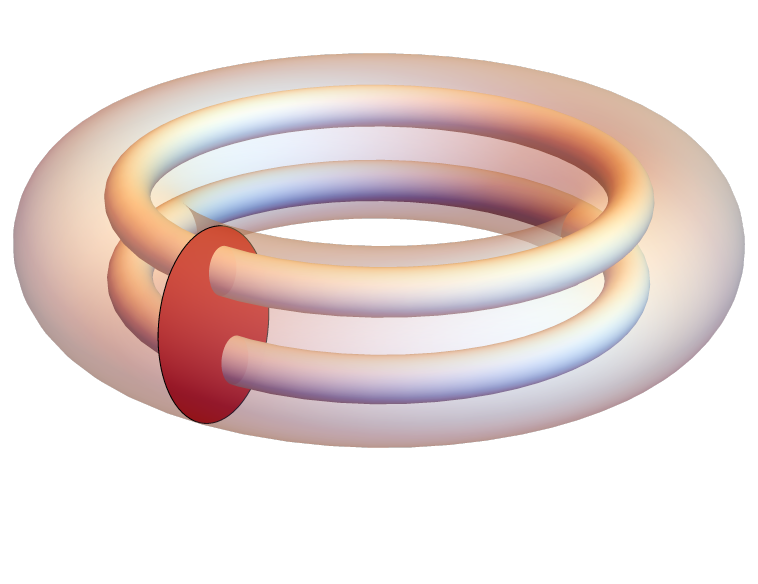}
            \put (31,38) {1}
            \put (32,27) {0}
            \put (-10,40) {\Huge{$\{$}}
        \end{overpic}
  \end{minipage}
  \begin{minipage}{.05\textwidth}
    \qquad \Large{$+$}
  \end{minipage}
    \begin{minipage}{.35\textwidth}
        \begin{overpic}[width=5cm]{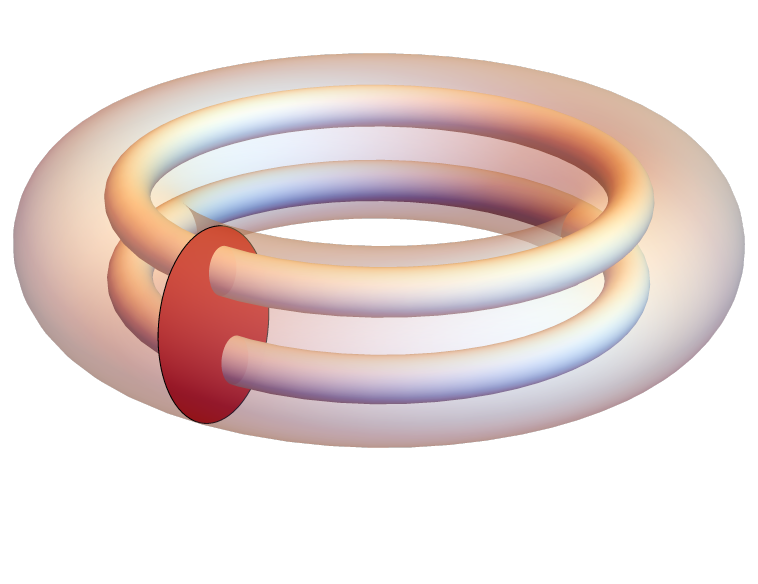}
            \put (31,38) {1}
            \put (32,27) {1}
            \put (100,40) {\Huge{$\}$}}
        \end{overpic}
  \end{minipage}
\caption{Topological realization of quantum Fourier transform $S$ on a state $\ket{a}$.}
\label{QFTSumPic}
\end{figure}

Acting on $\ket{a}$ with the modular $S$ operator on $\mathcal{H}_{\mathbb{T}^2}$ produces a superposition of basis states with coefficients given by the matrix elements $S_{ba}$ of Eq.\ \eqref{HadamardMatrix}. For $SU(2)_1$ at $a = 0$, applying Eq.\ \eqref{WnWilsonLoop} yields
\begin{equation}\label{WilsonQFT}
    S \ket{0} \;=\; \tfrac{1}{\sqrt{2}}\big(\ket{0} + \ket{1}\big) \;=\; \tfrac{1}{\sqrt{2}}\!\left[\,Z_\eta(R_0) + Z_\eta(R_1)\,\right],
\end{equation}
where $Z_\eta(R_a)$ denotes the path integral over $\eta$ with a Wilson loop along the core in the integrable representation $R_a$, and the equality is the path-integral realization of the linear combination of basis states. The analogous identity for $a = 1$ replaces the $+$ in Eq.\ \eqref{WilsonQFT} with a $-$. Applying these identities to the action of the Hadamard $S$ on the basis states $X\ket{0} = \ket{1}$ and $X\ket{1} = \ket{0}$ enables $SX\ket{0}$ and $SX\ket{1}$ to be expressed topologically as the sum over manifolds shown in Figure~\ref{QFTSumPic}.
From the above, we can interpret the action of the full Clifford group $\mathcal{C}_1$ on $W_2$ in terms of sums of topological tori, specifically
\begin{equation}
    \begin{split}
        S_0W_2 &= \frac{1}{\sqrt{2}}\left(\ket{00} -\ket{10}+\ket{01}+\ket{11} \right),\\
        S_1W_2 &= \frac{1}{\sqrt{2}}\left(\ket{00} -\ket{01}+\ket{10}+\ket{11} \right).
    \end{split}
\end{equation}

Constructing the action of $\mathcal{C}_n$, where $n>1$, requires the addition of the $C_{sum}(i,j)$ operator. We can express the action of $C_{sum}(i,j)$ on $W_2$ using the quantum Fourier transform $S$, for example
\begin{equation}\label{TwoToriC}
\begin{split}
    C_{sum}(1,2)W_2 
    &=\frac{i}{\sqrt{2}}\left(\ket{0} - \ket{1} \right) \otimes \ket{1},\\
    &= i\left(SX\ket{0} \right) \otimes \ket{1}.
\end{split}
\end{equation}
In the topological picture, the action of $C_{sum}(i,j)$ on $W_2$ corresponds to 
sum over tori, as in Figure \ref{QFTSumPic}, followed by the tensor product with an additional torus. where $X_2\ket{0} = \ket{1} = N_{a,0}^1 \ket{0}$. Therefore, from the action of $X_2$ we have
\begin{figure}[h]
  \begin{minipage}{.23\textwidth}
    \large{$C_{1,2}\ket{W_2}= \frac{1}{\sqrt{2}}$}
  \end{minipage}
  \begin{minipage}{.27\textwidth}
        \begin{overpic}[width=3.5cm]{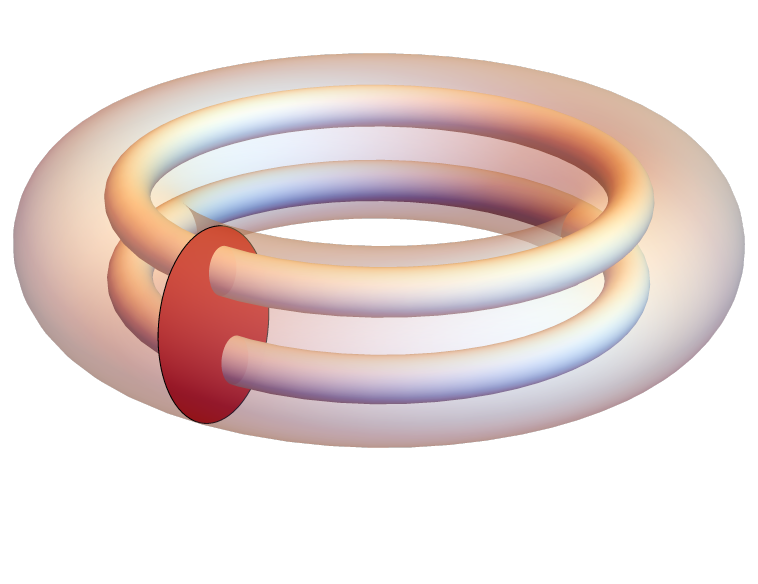}
            \put (31,38) {1}
            \put (32,27) {0}
            \put (-10,40) {\Huge{$\{$}}
        \end{overpic}
  \end{minipage}
  \begin{minipage}{.05\textwidth}
    \qquad \Large{$-$}
  \end{minipage}
    \begin{minipage}{.27\textwidth}
        \begin{overpic}[width=3.5cm]{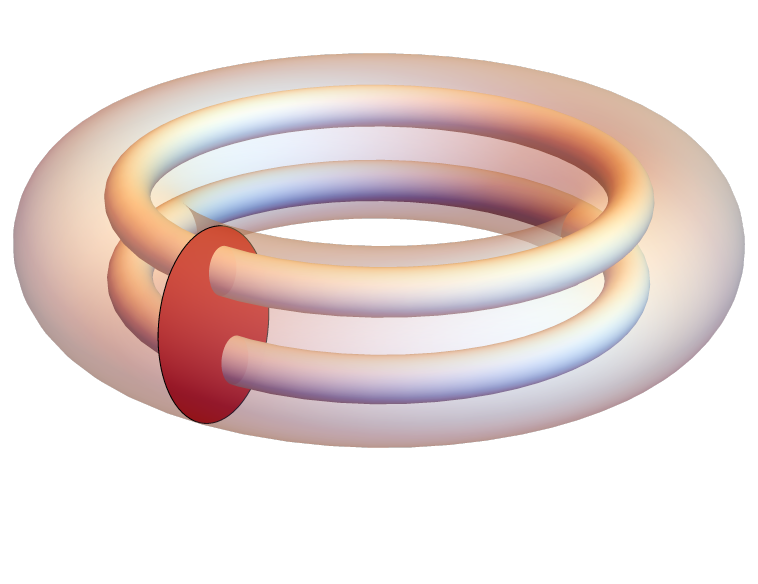}
            \put (31,38) {1}
            \put (32,27) {1}
            \put (100,40) {\Huge{$\}$}}
        \end{overpic}
  \end{minipage}
\begin{minipage}{.15\textwidth}
    \qquad \large{$\otimes X_2\ket{0}$}
  \end{minipage}
  \begin{minipage}{.23\textwidth}
    \large{$\qquad \qquad  = \frac{1}{\sqrt{2}}$}
  \end{minipage}
  \begin{minipage}{.27\textwidth}
        \begin{overpic}[width=3.5cm]{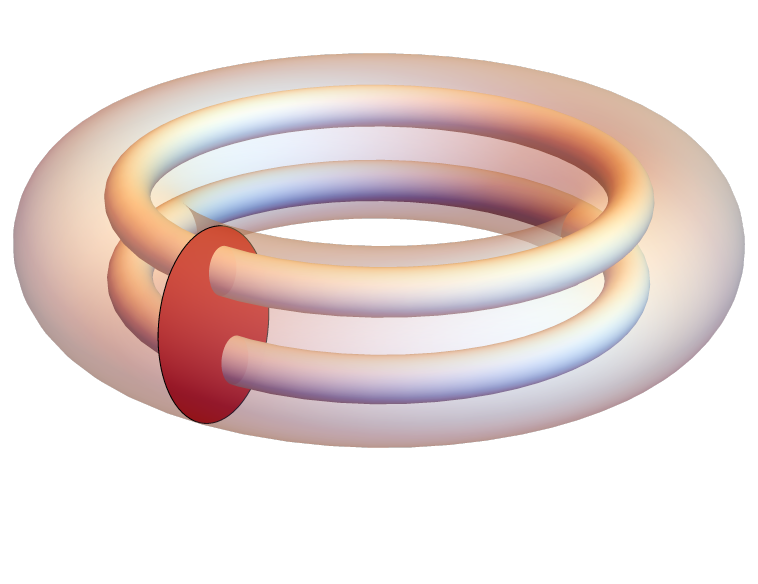}
            \put (31,38) {1}
            \put (32,27) {0}
            \put (-10,40) {\Huge{$\{$}}
        \end{overpic}
  \end{minipage}
  \begin{minipage}{.05\textwidth}
    \qquad \Large{$-$}
  \end{minipage}
    \begin{minipage}{.27\textwidth}
        \begin{overpic}[width=3.5cm]{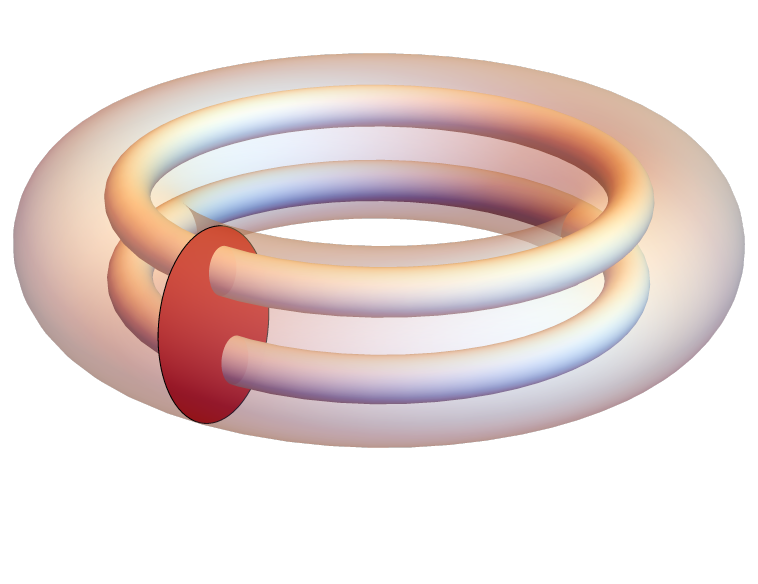}
            \put (31,38) {1}
            \put (32,27) {1}
            \put (100,40) {\Huge{$\}$}}
        \end{overpic}
  \end{minipage}\\
  \begin{minipage}{.50\textwidth} 
   $ $
  \end{minipage}
  \qquad \qquad \qquad
\begin{minipage}{.15\textwidth}
        \begin{overpic}[width=3.5cm]{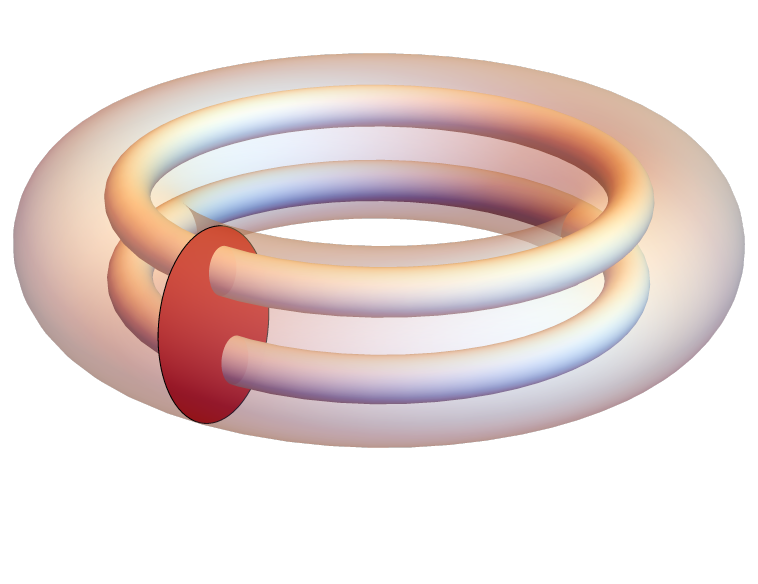}
            \put (31,38) {1}
            \put (32,27) {0}
            \put (-15,40) {\large{$\times$}}
        \end{overpic}
  \end{minipage}
\caption{Topological realization of the $C_{sum}$ gate acting on $W_2$.}
\label{CSumTopological}
\end{figure}
which is the sum of a product of $3$-dimensional tori. Recall the action of $\mathcal{C}_1$ on $W_2$ is also a sum of products of such topological tori. Computing  entropies involves sewing together $2$ tori as in Figure \ref{SumGraphRepOfPathInt}. More generally, Clifford operations on $W_2$ generate sums of products of tori, and computing entanglement entropies reduces to sewing together pairs of such manifolds, as in Figure~\ref{SumGraphRepOfPathInt}.

In topological quantum field theories, in particular $SU(2)_1$ Chern-Simons theory, the mapping class group naturally connects manifold manipulations to algebraic operations on the corresponding Hilbert spaces. For a surface $\Sigma_g$, of genus $g$, the mapping class group describes the different ways $\Sigma_g$ can be smoothly deformed into itself, up to a continuous equivalence. These deformations encode topological symmetries of $\Sigma_g$, and correspond to unitary transformations on the associated Hilbert space $\mathcal{H} \left( \Sigma_g \right)$ in $SU(2)_1$. In this way, these unitary transformations implement the modular action that relates the topology of $3$-manifolds with the algebraic structure of conformal blocks in the corresponding WZW model.

\begin{figure}[h]
\centering
\includegraphics[width=10cm]{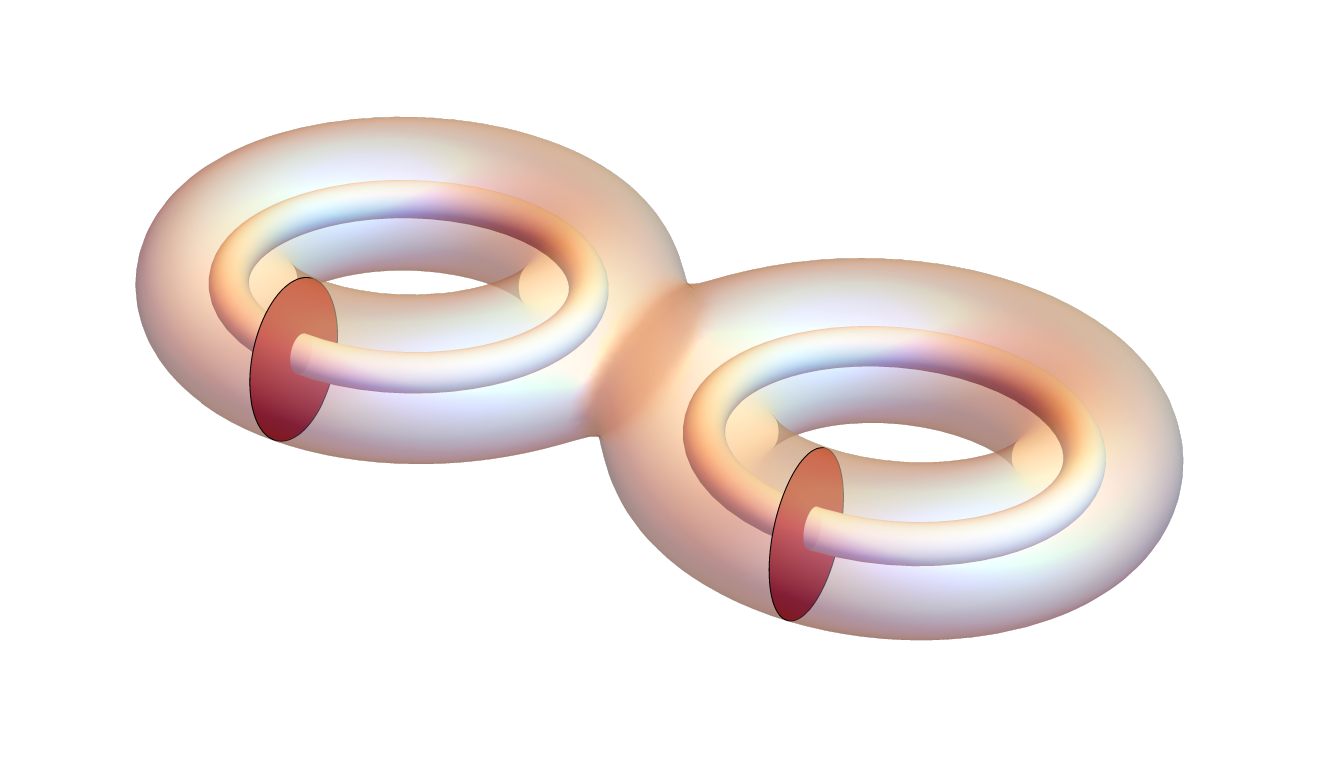}
\caption{Genus-$2$ handlebody $\Sigma_2$, with a solid torus removed from each genus-$1$ component. The manifold $\Sigma_2$, with a Dehn twist applied to each genus-$1$ component, describes the available unitary transformations on the associated Hilbert space. For states $W_2$ prepared on $\Sigma_2$, the diffeomorphisms define the Clifford orbit of $W_2$.}
\label{fig:DoubleTorus}
\end{figure}

For a genus-$1$ surface, the mapping class group is isomorphic to the modular group $SL(2,\mathbb{Z})$, generated by two elements $S$ and $T$ with matrix representations
\begin{equation}\label{TorusMatrices}
    S = \begin{bmatrix}
        0 & -1\\
        1 & 0
    \end{bmatrix}, 
    \quad
        T = \begin{bmatrix}
        1 & 1\\
        0 &1
    \end{bmatrix}. 
\end{equation}
The matrices in Eq.\ \eqref{TorusMatrices} correspond to the two diffeomorphisms of the torus, each a Dehn twist along the meridian and longitudinal curves of the torus, respectively. Geometrically, the Dehn twist transformation corresponds to cutting the torus along an incontractible cycle, rotating one side by $2\pi$, and gluing it back together. The matrices $S$ and $T$ generate the full modular group, and in $SU(2)_1$ form a Clifford group acting on the Hilbert space of a single anyon pair. In particular, the two solid tori appearing in $C_{i,j}W_2$, from Eq.\ \eqref{TwoToriC}, can be joined after Dehn twists to produce the action of $SX$, as shown in Figure \ref{fig:DoubleTorus}. This connection suggests the following conjecture.
\begin{conjecture}\label{Conj1}
    The Clifford group orbit of $W_2$ is represented by a genus-$2$ handlebody $\Sigma_2$, with two solid tori removed from the interior (see Figure \ref{fig:DoubleTorus}). A Dehn twist is implemented on each genus-$1$ component of $\Sigma_2$.
\end{conjecture}

Extending Conjecture\ref{Conj1} to consider Clifford action on $W_3$, we observe the following corollary.
\begin{corollary}\label{Corollary1}
    The Clifford group action on $W_3$ can be represented as a sum over genus-$2$ handlebodies, each constructed from two genus-$1$ components (as in Figure \ref{fig:DoubleTorus}), connected by Dehn twists generated by the modular $S$ and $T$ transformations.
\end{corollary}
A similar construction can be used to prepare Dicke states and their Clifford orbits. For example, the $4$-qubit Dicke state with two excitations is defined
\begin{equation}
    \ket{D_2^4} = \frac{1}{\sqrt{6}} \sum_{i\neq j}^{4} X_iX_j \ket{0000},
\end{equation}
where $X\ket{a} = N_{a,1}^{a+1}\ket{a, \mod 2}$. Following Corollary~\ref{Corollary1}, the state $\ket{D_2^4}$ and its Clifford orbit can be represented as a sum over $6$ genus-$2$ handlebodies. Similarly, $\ket{D_2^6}$ can be represented as a sum over $15$ genus-$2$ handlebodies, and $\ket{D_3^6}$ as a sum over $20$ genus-$3$ handlebodies. In general, the number of handlebodies required is specified by $\binom{n}{k}$, with $k$ the number of single-site excitations, which also determines the genus of the summand manifolds. 

\begin{figure}[h]
\centering
\includegraphics[width=10cm]{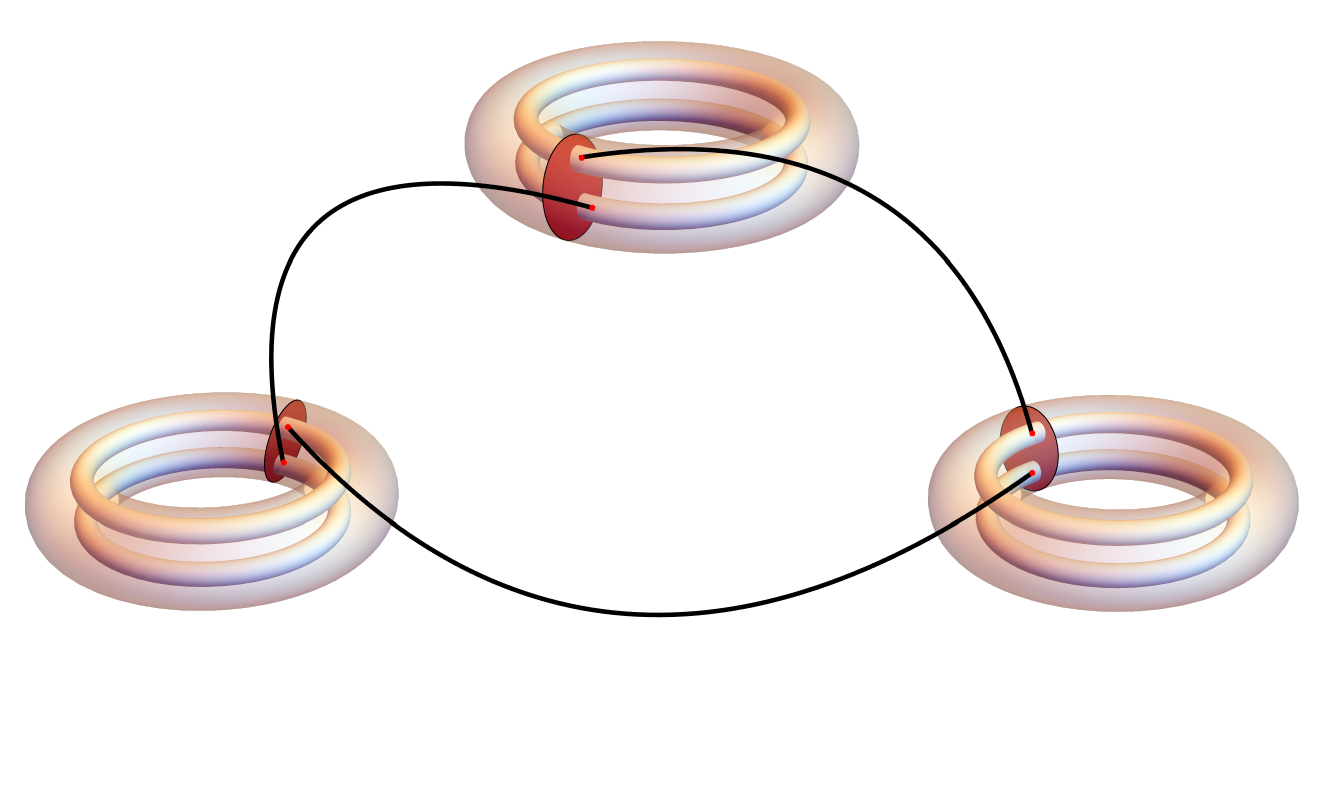}
\caption{Heegaard splitting of $3$-manifold into genus-$1$ handlebodies. Clifford operations can independently be realized on each of the constituent handlebodies, as defined in Eqs.\ \eqref{WilsonQFT} and \eqref{TwoToriC}.}
\label{fig:HeeygardSplitting}
\end{figure}

A decomposition of the $3$-manifolds in this section is provided by the concept of Heegaard splitting. For a $3$-manifold $M$ a Heegaard splitting writes $M$ as 
\begin{equation}
    M = \Sigma_1 \cup_f \Sigma_2,
\end{equation}
for $\Sigma_1$ and $\Sigma_2$ genus-$1$ and genus-$2$ handlebodies, respectively. The map $f: \partial \Sigma_1 \rightarrow \partial\Sigma_2$ provides a homeomorphism that glues the boundaries of the handlebodies together. For the genus-$1$ torus, distinct $3$-manifolds are generated by gluing together two solid tori using an element of the mapping class group of the torus boundary. Figure \ref{fig:HeeygardSplitting} illustrates the Heegaard splitting for the genus-$3$ handlebody used to prepare $W_3$.
The topology of the resulting $3$-manifold $M$ is determined by the gluing map $f$, with $S$ generating the manifold $S^3$, and a generic element of $SL(2,\mathbb{Z})$ generating a lens space. Accordingly for path integrals in $SU(2)_1$, the modular transformations $S$ and $T$ play the role of specifying the topological data of the manifold on which the theory is evaluated.

In this section we developed a topological framework representing Clifford group action in $SU(2)_1$. We demonstrated that the Fourier transform $S$ and the controlled-sum $C_{sum}$ operations admit representations as sums and products over the manifold $\eta$ introduced in Section \ref{Manifold}. The algebraic relations among the Clifford operators were shown to emerge as Wilson loop expectation values, providing a direct connection between the operator algebra and topological information of the manifold. At genus $g = 1$, where the mapping class group is the modular group $SL(2,\mathbb{Z})$, we explicitly identified the modular generators $S$ and $T$ with the corresponding single-qubit Clifford generators acting on $\mathcal{H}_{\mathbb{T}^2}$. We further proposed, in Conjecture~\ref{Conj1} for the Clifford orbit of $W_2$ and Corollary~\ref{Corollary1} for $W_3$, a higher-genus extension in which the Clifford orbit of $W$ and Dicke states is represented by a sum over genus-$g$ handlebodies with Dehn twists applied to each constituent torus. These higher-genus statements are conjectural and specific to $W$ and Dicke states. Additionally, we illustrated how Heegaard splittings decompose $3$-manifolds into handlebodies, glued along their boundaries, on which Clifford group actions can be realized. Finally, we remarked on the extension of this construction to Dicke states, as well as their respective Clifford orbits. We now conclude this work and present potential directions for future work in this area.

\section{Discussion}

In this work we develop a topological framework for preparing $W_n$ states in $SU(2)_1$ Chern-Simons theory, and computing their associated entanglement entropies. Building upon prior formulations of stabilizer states in $U(1)_k$, we extend the topological construction to non-stabilizer systems by interpreting the generalized $X$ operator as a fusion tensor in $SU(2)_1$. We construct this fusion tensor and its adjoint through path integration over a solid torus, with interior tori removed, giving a topological representation for $\ket{W_n}$. This approach not only reproduces $W_n$ density matrices, but also allows for a topological calculation of $W_n$ entanglement entropies as a sum over manifold replicas. We use this topological formulation to explicitly compute the entanglement entropies of $\ell$-party subsystems of $\ket{W_n}$. We conjecture an extension of this proposal for moving beyond $W_n$ states, to construct the full set of Dicke states.

Beyond preparing $W_n$ states, we extend the topological framework to include Clifford group dynamics by providing topological realizations of the algebraic Clifford generators (the quantum Fourier transform $S$, phase $P$, and controlled-sum $C_{sum}$ operators) within the $SU(2)_1$ Chern-Simons setting. Connecting the algebraic construction of Clifford operators, in the Kac-Moody algebra, to sums and products over tori, we establish a pathway to study Clifford orbits and entanglement evolution from a topological perspective. At genus $g = 1$, the modular group $SL(2,\mathbb{Z})$ acts on $\mathcal{H}_{\mathbb{T}^2}$ and the generators $S$ and $T$ are identified with the corresponding single-qubit Clifford generators, with each Clifford action represented by a Dehn twist on the boundary torus. For $W$ and Dicke states, we further conjecture (Conjecture~\ref{Conj1} and Corollary~\ref{Corollary1}) that the full Clifford orbit at $n \geq 2$ qubits is realized as a sum over genus-$g$ handlebodies with Dehn twists applied to each constituent torus. For states $\ket{W_n}$ prepared on $3$-manifolds, this formalism enables the topological analysis of entanglement dynamics under action of the Clifford group. Linking the algebra of quantum operators to the topological information of manifolds further provides a versatile toolkit for interpreting entanglement structure in quantum many-body systems, across distinct representations.

One natural extension of this work is to explore higher level Chern-Simons theories, i.e. $SU(2)_k$ with $k>1$ or even $SU(n)_k$, to determine how representations of the underlying algebra impact the entanglement structure of topological states. In the low-energy limit, the fractional quantum hall effect (FQHE) is effectively described by an $SU(n)_k$ Chern-Simons theory. While lower level $SU(2)_k$ theories correspond to Ising anyons, higher level generalizations allow for parafermionic excitations, a promising candidate for topological quantum computing~\cite{Wen1990, Fradkin1998}. These higher-level $SU(2)_k$ theories incorporate nontrivial fusion rules and braiding statistics derived from the modular $S$ and $T$ matrices, and enable fault-tolerant logical operations. A deeper understanding of the entanglement properties and operator dynamics within higher level Chern-Simons theories could offer new insights into the universality and robustness of topologically encoded quantum information.

An $SU(n)_k$ Chern-Simons theory defined on a $3$-manifold with genus-$2$ boundary admits a holographic dual description as a WZW conformal field theory (CFT) living on that boundary~\cite{Witten1989, Elitzur1989}. In this correspondence, the bulk topological degrees of freedom of the Chern-Simons theory project onto the chiral sectors of the boundary CFT, establishing a concrete realization of the bulk-boundary duality in a topological setting. States prepared topologically in $SU(n)_k$ correspond to conformal blocks of the boundary WZW model, while Wilson lines in the bulk map to primary operator insertions on the boundary~\cite{MooreSeiberg1989}. Consequently, topological manipulations, such as modular transformations or Dehn twists, in the $SU(2)_1$ Chern-Simons theory can be interpreted as generating nontrivial operator dynamics and entanglement evolution in the dual CFT. Conversely, related efforts have explored bulk dynamics generated by tuning the quantum properties of the boundary CFT~\cite{Dong:2016fnf,Dong:2018seb,Cao:2024nrx}. This correspondence provides a geometric and topological framework for understanding how entanglement dynamics in topological field theories manifest as information flow and operator evolution in associated holographic duals~\cite{Gukov2013,Ammon2013,Maloney2010}.

Finally, we seek to connect the protocols developed in this framework with practical quantum computing architectures. Efficient generation and distribution of precisely structured entanglement remains a challenge in contemporary quantum hardware and near-term quantum networks~\cite{Hein2003,Negrin:2024tyj,Hughes:2025jwa}. Moreover, both $W_n$ and Dicke states feature prominently in such applications~\cite{Prevedel2009,Chiuri2012,Wang2016}, and the well-characterized entanglement features of Dicke states~\cite{nepomechie2023qudit,Munizzi:2023ihc} enable enhanced quantum sensing capabilities~\cite{Chin:2024ctm}, as well as resource-efficient state preparation~\cite{Farhi2014,Zhou2018,Basso2022}. Given their extensive applications in quantum networks~\cite{Prevedel2009,Chiuri2012,Wang2016}, metrology, and error detection, a topological description of $W_n$ and Dicke state entanglement could inform more resource-efficient state preparation, and eventual fault-tolerant protocols, particularly in the context of topological quantum computing. Moreover, investigating whether Clifford dynamics realized through topological transformations can be experimentally implemented on quantum simulators or topological qubits would strength the connection between theory and physical implementation. In this way, the topological perspective developed here can provide conceptual clarity and potential design guidance for future quantum technologies.

%\begin{appendix}
%\addtocontents{toc}{\protect\setcounter{tocdepth}{1}}

%\end{appendix}

\bibliographystyle{JHEP}
\bibliography{ChernSimons}

\end{document}